\documentclass[aps,prl,reprint,superscriptaddress]{revtex4-2}

\usepackage{graphicx}
\usepackage{xcolor}
\usepackage{amsmath,esint}
\usepackage{amssymb,bm}
\usepackage{braket}
\usepackage{enumitem}
\usepackage{float}
\usepackage{tikz}
\usepackage[colorlinks=true,linkcolor=blue,citecolor=red]{hyperref}
\usepackage{mathtools}
\usepackage{tcolorbox}
\usepackage{relsize}
\usepackage{multirow}

\newcommand{\Li}{\mathcal{L}}

\newcommand{\rmd}{\text{d}}

\newcommand{\DelR}{\vec{\Delta}_{\text{r}}}
\newcommand{\DelT}{\Delta_{\text{t}}}

\newcommand{\Gs}{\Gamma_{\textsc{s}}}

\newcommand{\eref}[1]{(\ref{#1})}

\newcommand{\kvec}{\vec{k}_{1}}
\graphicspath{{./figures/}}

\newcommand{\rhois}{\rho_{\textsc{is}}}
\newcommand{\rhoist}{\rho_{\textsc{ist}}}
\newcommand{\rhoes}{\rho_{\textsc{es}}}
\newcommand{\rhoest}{\rho_{\textsc{est}}}
\newcommand{\fsr}{f_{\textsc{lrt}}}
\newcommand{\rat}{\kappa}

\newcommand{\be}{\begin{equation}}
\newcommand{\ee}{\end{equation}}
\newcommand{\ben}{\begin{eqnarray}}
\newcommand{\een}{\end{eqnarray}}
\newcommand{\bF}{\begin{figure}}
\newcommand{\eF}{\end{figure}}
\newcommand{\bi}{\begin{itemize}}
\newcommand{\ei}{\end{itemize}}

\newcommand{\defeq}{\stackrel{\text{def}}{=}}

\usepgfmodule{nonlineartransformations}
\tikzset{lasernode/.style={
    top color=red,
    bottom color=red,
    middle color = white
    }
}

\usetikzlibrary{decorations.pathmorphing}

\tikzset{snake it/.style={decorate, decoration=snake}}

\begin{document}

\title{Signatures of Correlation of Spacetime Fluctuations in Laser Interferometers}

\author{B.~Sharmila}
\email{Sharmila.Balamurugan@warwick.ac.uk}
\affiliation{Department of Physics, University of Warwick, Coventry CV4 7AL, UK.}

\author{Sander M. Vermeulen}
\email{smv@caltech.edu}
\affiliation{Division of Physics, Mathematics and Astronomy,
California Institute of Technology, Pasadena, CA 91125, USA.}

\author{Animesh Datta}
\email{animesh.datta@warwick.ac.uk}
\affiliation{Department of Physics, University of Warwick, Coventry CV4 7AL, UK.}

\date{\today}

\begin{abstract}
Spacetime fluctuations (SFs), a common feature of different proposed gravity models, could be detected using laser interferometers. In the search for SFs, a correspondence between the expected output signals and different gravity models is needed,  both for guiding the design of future interferometers, and for identifying the signal in experimental data. In this work, we provide such a correspondence for some classes of SFs and geometries of the interferometers. We consider three different classes of SFs, characterised by the decay behaviours and symmetries of their two-point correlation functions. Our approach applies to Michelson laser interferometers with Fabry-P\'erot arm cavities such as the km-long LIGO detectors and those without arm cavities such as the laboratory-scale setups QUEST and GQuEST. Analysing the expected interferometer output signals, we identify three characteristic signatures for each class of SF. The designed broadband sensitivity of the laboratory-scale instruments would allow all characteristic signatures of the different classes of SFs to be observed, and such observations could provide more information on the nature of the SFs than those from LIGO. On the other hand, we find that LIGO is better suited for detecting the bare presence or absence of SFs.
\end{abstract}



\maketitle


\section{Introduction}

A wide spectrum of ideas have been considered~\cite{GravHist,Bassi2017} to understand the fundamental nature of gravity;
some, such as the idea of spacetime fluctuations (SFs), form a \textit{leitmotif} in this effort. Since its first proposal by Wheeler~\cite{wheeler1957}, SFs have been extensively examined~\cite{Carlip2023} in the context of a quantum description of gravity, as well as in different semiclassical models of gravity~\cite{MesoGrav2024,Oppenheim2023} and the study of stochastic gravitational waves~\cite{stochGW1,stochGW2}. Spacetime has thus been hypothesised to be, for instance, classical but stochastic~\cite{OppenPQG}, or classical but emerging from underlying quantum entanglement~\cite{EntangST}, or holographic and having quantum perturbations~\cite{geontropy2023}. 
These hypotheses suggest different mathematical forms for the correlation functions of SFs.

The scale of the correlations in SFs differs widely across models, ranging from the Planck length scale in effective field theories~\cite{Carney2024} to long-range correlations in holographic models~\cite{VerZurek21,kwon2025}.  The latter hypothesise observable effects such as the violation of Lorentz invariance~\cite{LIviolate}, gravitational decoherence~\cite{Oniga2016,Bassi2017}, blurring of astronomical objects~\cite{Lee2024}, and interferometric noise~\cite{amelino99,Shar2024,smv24}. Of these, interferometric noise has garnered much attention through the development of gravitational-wave detectors~\cite{GroteGEO2010}, and the Holometer~\cite{HoloData,Hogan2008} experiment. More recently, sophisticated laboratory-scale Michelson laser interferometers (MLIs), such as QUEST~\cite{Sander2021} and GQuEST~\cite{smv24}, which incorporate new quantum technologies, aim to search for SFs. 

Detecting SFs would constitute a breakthrough in understanding the fundamental nature of gravity. To enable such a detection, experimental designs require estimates of the the strength and bandwidth of the expected interferometric output signal so they can be suitably optimised. Further, computing the interferometric output signal can help in understanding how SF signals, which are typically broadband, can be distinguished from instrumental noise. Moreover, quantitative theoretical predictions of the output signal are essential either to rule out the presence of SFs in interferometric data or, if the underlying gravity theory includes free parameters, to constrain those parameters. While there are many theoretical works that have modelled the interferometric output signal to identify possible SF signatures~\cite{QLtIRB1,QLtIRB2,geontropy2023,Oppenheim2023,gardner2024,Carney2024}, they only offer model-specific predictions. 

We take a more expansive approach in studying SFs.
Irrespective of the classical or quantum description of a phenomenon,
correlations in physical processes tend to decay either exponentially or polynomially with increasing separation between two spacetime points~\cite{AshcroftMermin,NoneqStat}.
Exponential decay of correlations typically emanates from underlying physics that is short-ranged.
In the context of gravity, this captures both quantum~\cite{EntangST} and semiclassical~\cite{MesoGrav2024} models.
On the other hand, polynomial decay of correlations of the form $r^{-\eta}$ ($\eta\in\mathbb{R}$ and $\eta>0$), corresponds to long-range interactions.
In the context of gravity, such correlations of SFs are more commonly expected to decay as a reciprocal of distance, i.e., $1/r$~\cite{Karolyhazy66,geontropy2023,Lee2024,Carney2024} than with a higher inverse power-law such as $r^{-4/3}$~\cite{figurato2024}. 
Furthermore, internal symmetries of the spacetime metric are reflected in induced symmetries of the correlation function, such as factorisation into spatial and temporal parts. This factorisation occurs in SF models such as the Oppenheim model~\cite{Oppenheim2023}, a generalised K$\acute{\text{a}}$rolyh$\acute{\text{a}}$zy model~\cite{figurato2024}, the continuous spontaneous localization model~\cite{Ghirardi1990}, and the Di$\acute{\text{o}}$si-Penrose model~\cite{Bassi2017}. 

This motivates us to consider three possible classes of two-point correlation functions of the SFs:
(a) factorised into spatial and temporal correlations, both decreasing with increasing spatial and temporal separation (for instance, a semiclassical model~\cite{Oppenheim2023});
(b) an inverse of the separation between the two spacetime points (for instance, an SF model obeying the wave-equation in 3+1 dimensions~\cite{Karolyhazy66,geontropy2023,Lee2024}); and 
(c) an exponential decay with the separation between the two spacetime points (for instance, due to quantum entanglement~\cite{EntangST}, and in semiclassical~\cite{MesoGrav2024} models). 
Finding experimental evidence for or against one of the above classes of correlation functions describing SFs would constitute a breakthrough in elucidating the fundamental nature of gravity. 

In this work, corresponding to each class of SFs, we identify three characteristic signatures: the low-frequency behaviour, the high-frequency behaviour and the $\Li$ dependence of the interferometric output signal. Here $\Li$ is the arm-length of the MLI. In order to observe all three signatures, it is sufficient if the MLIs are sensitive over two decades spanning the light-round-trip frequency $\fsr=c/(2 \Li)$. However, even MLIs that are sensitive in a shorter span that includes $\fsr$, could allow observation of the three signatures to a limited extent. The projected sensitivities of laboratory-scale MLIs such as QUEST and GQuEST span the $\fsr$, unlike LIGO. Therefore, we find that these laboratory-scale MLIs allow observation of more signatures than LIGO in principle, thus providing more information on the class of the underlying correlation function and aiding in distinguishing between the classes (a)-(c).  
 
We also show that MLIs with arm cavities have a significant advantage in detecting the bare presence or absence of SFs. This is due to the peak in the interferometric output signal for MLIs with arm cavities at their $\fsr$. We note that for LIGO, $\fsr\approx 37.5$~kHz, which is outside the frequency span of the publicly available data~\cite{LIGOunpub}. 

Our approach is agnostic to the microscopic origins of the SFs~\footnote{They differ widely between the different gravity models, ranging from quantum entanglement of degrees of freedom resulting in suitable moments of the metric~\cite{EntangST} to \textit{geontropic}~\cite{geontropy2023} or stochastic~\cite{MesoGrav2024} fluctuations of the metric itself.}.
It requires as input only the two-point correlation of the spacetime metric from any model of gravity and the geometry of the MLI. 
This enables us to compute the output signal power spectral density (PSD) of the MLI on which all our conclusions are based. Our approach also accommodates computing the signal PSD for MLIs with arm cavities, such as LIGO. Whether the presence of arm cavities provides any advantage towards detection has been much-debated lately~\cite{kwon2016,VerZurek21,geontropy2023}. This debate even lead to the exclusion of arm cavities from experimental designs of recent MLIs~\cite{HoloData,Sander2021,smv24}.
We now settle the debate for any SF from classes (a)-(c). 

\section{Methodology}
\subsection{Modelling light propagation} 
To investigate the effects of a fluctuating spacetime, we consider an isotropic spacetime metric $g^{\beta \gamma}$ ($\beta,\gamma=0,1,2,3$) of the form 
\begin{equation}
g^{00}=-1 + 2 w(\bm{r}), \quad g^{ij}=\delta_{ij}, \quad g^{0i}=g^{i0}=0,
\end{equation}
where $i,j=1,2,3$ and $w(\bm{r})$ is a random process in $\bm{r}\equiv (t,x,y,z)$, with $w \ll 1$. Here 0 (resp., $i\in\{1,2,3\}$) corresponds to the timelike (resp., spacelike) component. While we consider this specific form in this work, our approach can 
encompass general fluctuations in every $g^{\beta \gamma}$ (see Supplementary Material~\cite[Sec. I]{supp} for details). To model the propagation of light of frequency $\Omega$ and wavelength $\lambda=2\pi c/\Omega$ in the given spacetime manifold, we solve for the electromagnetic tensor in the relativistic wave equation (RWE) (see~\cite[Sec. I]{supp} with Ref. \cite{Tsagas2005} therein for details).
This is subject to the following assumptions.

\begin{enumerate}[wide, labelwidth=!, labelindent=0pt,label=\textbf{Assumption (\roman*)}]
\item \label{Ass:eik} \textit{Setting length and time scales:} The correlation scales in length and time of the metric fluctuations $w$ need to be longer than $\lambda$ and $2\pi/\Omega$ respectively. This effectively sets the wavelength as the smallest length scale in the system, i.e., the \textit{eikonal approximation}. This allows us to neglect diffraction due to the fluctuations.
\end{enumerate}
Applying the \textit{eikonal approximation} ($k=2\pi/\lambda \to \infty$), the RWE reduces to the light propagating along the null geodesic for any general spacetime metric~\cite{PaddyGrav}.
For light propagating along, say, the $z$-axis, the electric field solution for the RWE (see~\cite[Sec. I]{supp} for details) is
\begin{eqnarray}
\vec{E}(\bm{r}(t))=\vec{E}_{in}(x,y) e^{i k \Phi(\bm{r}(t))},
\label{eqn:Soln}
\end{eqnarray}
where
\begin{subequations}
\begin{align}
&\Phi(\bm{r}(t))=c t -z + \Phi_{\textsc{f}}(\bm{r}(t)),\\
&\Phi_{\textsc{f}}(\bm{r}(t)) = c  \int\limits_{0}^{t} \text{d}t' \, w(\bm{r}(t')),
\end{align}
\label{eqn:APhi}
\end{subequations}
and $\vec{E}_{in}(x,y)$ is the input transverse profile of the beam. 

\begin{enumerate}[wide, labelwidth=!, labelindent=0pt,label=\textbf{Assumption (\roman*)}]
\setcounter{enumi}{1}
\item \label{Ass:SVEA} \textit{Slowly varying envelope approximation (SVEA):} 
The above solution \eref{eqn:Soln} also uses the SVEA which is consistent with the eikonal approximation.
It assumes a very small rate of metric-fluctuation-induced phase fluctuations $\partial_{t} \Phi_{\textsc{f}} \ll c$ and $\partial_{i} \Phi_{\textsc{f}}\ll 1$ ($i=x,y,z$).
\end{enumerate}

We now introduce the assumptions on the random SF process $w$.
\begin{enumerate}[wide, labelwidth=!, labelindent=0pt,label=\textbf{Assumption (\roman*)}]
\setcounter{enumi}{2}
\item \label{Ass:stat} \textit{Stationarity:} $w(\bm{r})$ is a stationary Gaussian random process with the expectation values,
\begin{align}
&\overline{w}=0, \quad \text{and} \\ 
&\overline{w(t_{1},\vec{r}_{1}) w(t_{2},\vec{r}_{2})}  = \Gs \, \rho\left( c t_{12}, \vec{r}_{12} \right). \label{eqn:Corr_defn}
\end{align}
Here $\textsc{x}_{12}=\textsc{x}_{1}-\textsc{x}_{2}$ ($\textsc{x}=t,\vec{r}$) with $\vec{r}_{i}\equiv\{x_{i},y_{i},z_{i}\}$ ($i=1,2$) and $\Gs$ is  the strength of the correlation function. 
Both $\Gs$ and $\rho$ are dimensionless, consistent with $w$ being dimensionless.
Subsequently, we consider the correlation function $\rho\left( c t_{12}, \vec{r}_{12} \right)$ to be from the classes (a)-(c) to obtain the corresponding MLI output as illustrated in Fig.~\ref{fig:Corr}. 

\item \label{Ass:iso} \textit{Isotropy:} The two-point correlation function $\rho$ is isotropic in space, i.e., 
\begin{align}
\nonumber &\rho\left(\delta_{1}, \{ \delta_{2}, \delta_{3}, \delta_{4}\} \right) = \rho\left(\delta_{1}, \{ \delta_{4}, \delta_{2}, \delta_{3} \} \right)\\
 &= \rho\left(\delta_{1}, \{ \delta_{4}, \delta_{3}, \delta_{2}\} \right) = \dots,
\end{align}
for any separation $\delta_{i}$ ($i=1,2,3,4$). We consider the correlation length of $\rho$ to be $\ell_{r}$ for all three spatial dimensions, and the correlation time to be $\ell_{r}/c$. 
\end{enumerate}

\begin{figure}
\begin{tikzpicture}
\draw[red,thick] (0.,0.05) -- (1.95,0.05);
\draw[red,thick] (2.,0.) -- (4.5,0.);
\draw[red,thick] (2.,-2.) -- (2.,0.);
\draw[red,thick] (1.95,0.05) -- (1.95,2.);
\draw[red,thick,->] (0.,0.05) -- (1.,0.05);
\draw[red,thick,->] (2.,0.) -- (3.,0.);
\draw[red,thick,->] (1.95,0.05) -- (1.5,0.05);
\draw[red,thick,->] (4.,0.) -- (3.5,0.);
\draw[red,thick,->] (2.,0.) -- (2.,-1.5);
\draw[red,thick,->] (1.95,0.05) -- (1.95,1.);
\draw[red,thick,->] (1.95,2.) -- (1.95,1.5);
\draw[black,thick] (4.5,-0.2) -- (4.5,0.2);
\draw[black,thick] (1.75,2.) -- (2.15,2.);
\draw[black,thick] (1.85,-0.2) node[left] (scrip) {BS};
\draw[black!30,fill=black!30] (1.8,-0.2) -- (2.2,0.2)--(2.15,0.25)--(1.75,-0.15)--cycle;
\draw[black,thick] (1.8,-0.2) -- (2.2,0.2);
\draw[black,thick,<-] (2.1,-0.5) -- (2.8,-0.5) node[right] (scrip) {$\Li+\frac{\varphi_{\text{off}}}{k}$};
\draw[black,thick,->] (4.2,-0.5) -- (4.5,-0.5);
\draw[black,thick,<-] (1.5,0.1) -- (1.5,0.8) node[left] (scrip) {$\Li$};
\draw[black,thick,->] (1.5,0.7) -- (1.5,1.9);
\draw[black,fill=black!30] (-0.1,-0.1) rectangle (0.5,0.2);
\draw[lasernode] (-0.05,-0.05) rectangle (0.45,0.15);
\draw[black,thick,fill=red] (2.2,-1.8) -- (1.8,-1.8) arc(180:360:0.2) --cycle;
\draw[black,thick,snake it] (2,-2) -- (2.1,-2.5);
\draw[black,thick] (0.,0.05) node[left] (scrip) {A};
\draw[black,thick] (1.9,-2.) node[left] (scrip) {B};
\draw[black,thick] (4.5,0.) node[right] (scrip) {D};
\draw[black,thick] (2.,2.2) node[right] (scrip) {C};
\draw[black,thick,->] (0.2,-1.8) -- (0.5,-1.8) node[right] (scrip) {z};
\draw[black,thick,->] (0.2,-1.8) -- (0.2,-1.5) node[left] (scrip) {x};
\draw[black,thick,->] (0.2,-1.8) -- (0.1,-2) node[left] (scrip) {y};
\draw[black,thick] (-0.5,2.) node[left] (scrip) {(a)};
\end{tikzpicture}
\begin{tikzpicture}
\draw[black,thick] (1.,0.8) node[right] (scrip) {(b)};
\draw[red,thick] (2.,0.) -- (6.,0.);
\draw[red,thick] (2.,0.) -- (2.,-0.4);
\draw[red,thick,->] (6.,0.) -- (5.,0.);
\draw[red,thick,->] (2.,0.) -- (4.5,0.);
\draw[black,thick] (6.,-0.4) -- (6.,0.4);
\draw[black,thick] (1.65,-0.4) node[left] (scrip) {BS};
\draw[black!30,fill=black!30] (1.6,-0.4) -- (2.4,0.4)--(2.3,0.5)--(1.5,-0.3)--cycle;
\draw[black,thick] (1.6,-0.4) -- (2.4,0.4);
\draw[black, thick] (3.5,-0.4) node[below] (scrip) {M} -- (3.5,0.4);
\draw[green!40!black,thick,<-] (2.9,0.6)node[above] (scrip) {$T_{\textsc{m}}$} -- (3.4,0.6);
\draw[green!40!black,thick,->] (3.5,0.6) -- (4.,0.6) node[above] (scrip) {$R_{\textsc{m}}$};
\draw[black,thick,<->] (2.1,-1) -- (3.4,-1);
\draw[black,thick] (2.5,-1.2) node[right] (scrip) {$d_{\textsc{d}}$};
\draw[black,thick,<-] (3.6,-1) -- (4.5,-1) node[right] (scrip) {$\Li$};
\draw[black,thick,->] (5,-1) -- (5.9,-1);
\draw[black,thick] (6.,0.) node[right] (scrip) {D};
\end{tikzpicture}
\caption{(a) A schematic diagram of the a Michelson laser interferometer (MLI). Each of the two MLIs in Holometer~\cite{HoloData}, QUEST~\cite{Sander2021}, and GQuEST~\cite{smv24} is an MLI. (b) Fabry-P\'erot arm cavity design used in LIGO-type interferometers. Mirror M is introduced in each of the two arms of the MLI to create optical resonators (illustrated for arm D).
}
\label{fig:HoloSetup}
\end{figure}
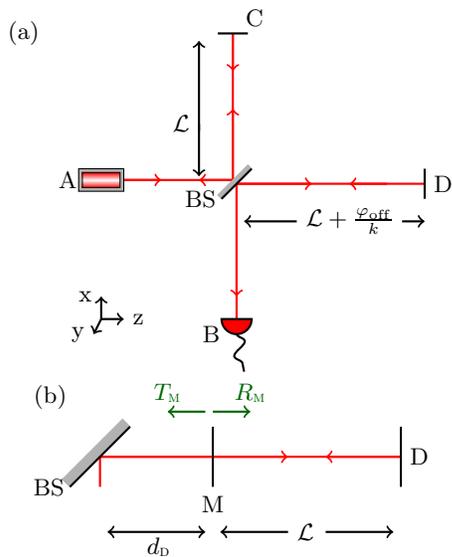

\subsection{Correlation classes}
We now list the different correlation classes, which completes the description of the spacetime fluctuations.

\noindent (a) {\it Factorised correlation function:} This class covers
correlation functions of the form ${\rho_{\textsc{f}}( c \DelT, \DelR )=\rho_{\text{s}}(\DelR) \rho_{\text{t}}(c \DelT)}$ where $\DelR$ is any 3-vector in space and $\DelT$ is any time interval. Gravity models that predict this class of correlations are the continuous spontaneous localization model~\cite{Ghirardi1990}, the Di$\acute{\text{o}}$si-Penrose model~\cite{Bassi2017} and the Oppenheim model~\cite{Oppenheim2023}. We concentrate here specifically on the Oppenheim model~\cite{Oppenheim2023} where
\begin{align}
\rho_{\textsc{f}}\left( c \DelT, \DelR \right)= - \left(\frac{\Vert \DelR \Vert}{\ell_{r}}\right) \: \left(\frac{\ell_{r} \,\delta(\DelT)}{c}\right).
\label{eqn:FactCorrExpr}
\end{align}
We note that multiplicative factors consistent with \ref{Ass:iso} are suitably introduced such that the delta-correlated $\rho_{\textsc{f}}$ remains dimensionless. Also, it is evident that $\rho_{\textsc{f}}$ is independent of $\ell_{r}$. Thus, the only parameter in this correlation class is the correlation strength~$\Gs$ that scales the correlation function (Eq.~\eref{eqn:Corr_defn}).

\vspace*{1 ex}
\noindent (b) {\it Inverse correlation functions:} This class, a subset of the polynomial decay of the correlations, is motivated by models such as those of Karolyhazy~\cite{Karolyhazy66}, and Zurek~\cite{geontropy2023}, as well as effective field theories~\cite{Carney2024}. 
We consider two  sub-classes where the correlations decay as a function of spatial separation and spacetime separation, respectively. 
\begin{enumerate}[label=(b\arabic*)]

\item \label{classB1} Spatial separation: 
\begin{align}
\rhois\left( c \DelT, \DelR \right)= \frac{\ell_{r}}{\Vert \DelR \Vert} \Theta(\Vert \DelR \Vert - c |\DelT|)
\end{align}
Such correlation is found in models that assume the fluctuations satisfy the wave equation~\cite{Karolyhazy66,geontropy2023,Carney2024}.

\item \label{classB2} Spacetime separation~\footnote{Carrying forward the step function from \ref{classB1}, the sign convention in the spacetime separation is chosen such that the correlation remains a real-valued function.}: 
\begin{align}
\rhoist\left( c \DelT, \DelR \right)= \frac{\ell_{r} \: \Theta(\Vert \DelR \Vert - c |\DelT|) }{\sqrt{\Vert \DelR \Vert^{2} - c^{2} \DelT^{2}}}.
\end{align}
This is a generalisation  of \ref{classB1}. 
\end{enumerate}
While we use two parameters to characterise strength $\Gs$ and the scale $\ell_{r}$ in this class, the function depends only on the product $\Gs \ell_{r}$, rendering the parameters degenerate.

\vspace*{1 ex}
\noindent (c) {\it Exponential correlation functions:} This class of correlation functions covers models motivated by entanglement between holographic degrees of freedom~\cite{EntangST} or a mesoscopic interpretation of gravity~\cite{MesoGrav2024}. 
We again consider two  sub-classes based on 
\begin{enumerate}[label=(c\arabic*)]

\item Spatial separation: 
\begin{align}
\rhoes\left( c \DelT, \DelR \right)= e^{-\frac{\Vert \DelR \Vert}{\ell_{r}}} \Theta(\Vert \DelR \Vert - c |\DelT|).
\end{align}

\item Spacetime separation: 
\begin{align}
\rhoest\left( c \DelT, \DelR \right)= e^{-\frac{\sqrt{\Vert \DelR \Vert^{2} - c^{2} \DelT^{2}}}{\ell_{r}} } \: \Theta(\Vert \DelR \Vert - c |\DelT|).
\end{align}
\end{enumerate}
Class (c) is a true two-parameter model, unlike classes (a) and (b). Classes (b) and (c) cannot be factorised into spatial and temporal functions.

\subsection{MLI output signal PSDs}
In an MLI (Fig. \ref{fig:HoloSetup} (a)), light propagates from a laser source at the input port A to a detector at the output port B via the two perpendicular arms, denoted by C and D. The 50/50 lossless beamsplitter~\footnote{The beamsplitter and the end mirrors are assumed to be perfect due to the very small optical losses that are typical in these high-precision interferomters.} is denoted by BS and is taken as the origin of the reference frame in our computation. We can effectively assume the detector to be at the origin, as the effect of any phase fluctuations suffered by the light after interference at the BS is negligible to that of phase fluctuations incurred in the arms in practice. The arm length of the MLI without fluctuations is $\Li$ and $\tau_{0}=1/\fsr$ (we recall that $(\fsr)^{-1}=2\Li/c$). 
The PSD of the optical path difference between the two arms (see~\cite[Sec. II]{supp} for details) at the detector is then written as a cosine transform from time separation $\Delta_{\tau}$ to frequency $f$.
\begin{align}
S(f)= \frac{c^{2} \Gs }{2 \pi} & \int_{0}^{\infty} \rmd \Delta_{\tau}  \Bigg[\sigma(\Delta_{\tau}) - \xi(\Delta_{\tau}) \bigg] \cos 2 \pi f \Delta_{\tau},
\label{eqn:PSDholo}
\end{align}
where
\begin{subequations}
\begin{eqnarray}
\nonumber &&\sigma(\Delta_{\tau})= \int\limits_{0}^{\tau_{0}} \rmd t_{1} \int\limits_{0}^{\tau_{0}} \rmd t_{2} \, \rho( c (t_{1}+\Delta_{\tau}-t_{2}),0,0,s(t_{1})-s(t_{2}))\\
&& = \int\limits_{0}^{\tau_{0}} \rmd t_{1} \int\limits_{0}^{\tau_{0}} \rmd t_{2} \, \rho(c(t_{1}+\Delta_{\tau}-t_{2}),s(t_{1})-s(t_{2}),0,0),
\label{eqn:SigDefn}\\
\nonumber &&\xi(\Delta_{\tau}) = \int\limits_{0}^{\tau_{0}} \rmd t_{1} \, \int\limits_{0}^{\tau_{0}} \rmd t_{2} \, \\
&& \hspace*{5 em} \rho(c(t_{1}+\Delta_{\tau}-t_{2}),s(t_{1}),0,-s(t_{2})).\label{eqn:XiDefn}
\end{eqnarray}
\label{eqn:CorrSet}
\end{subequations}
Also, $s(t)=c t$ if $t \leqslant \tau_{0}/2$ and $s(t)=2 \Li - c t$ if $t>\tau_{0}/2$. Assumptions (i)-(iv) are used to obtain Eq. \eref{eqn:PSDholo}, with Eq. \eref{eqn:SigDefn} using \ref{Ass:iso}. We reiterate that the expressions obtained in this subsection are true irrespective of the form of the correlation function $\rho$ and are not limited to classes (a)-(c). 
$\sigma$ arises from correlations in the spacetime metric fluctuations within an arm of the interferometer, and $\xi$ corresponds to correlations of the metric fluctuations between the two arms. Here, we have also assumed that the width of the light beams are of the order of the wavelength and negligible (see~\cite[Sec. II]{supp} for details)

A response-function-based approach \footnote{Both our approaches of obtaining the spectral densities, namely, based on correlation integral in Eq.~\eref{eqn:PSDholo} and the response function in Eq.~\eref{eqn:PSDirf} assume weak stationarity of the optical path difference. However, the existence of such a stationarity is not guaranteed, even when assuming the underlying SFs to be stationary and Gaussian. Therefore, when computing the PSDs, we check if the covariance of the optical path difference is positive-definite, to ensure weak stationarity of the optical path difference in our investigation. We find in class (c), this holds only when $\rat=\ell_{r}/\Li\ll1$.} allows straightforward extension of our methodology to different MLI geometries and detection schemes. 
These include for instance, the cross spectral density (CSD) of the output signal from two co-located, co-aligned MLIs (see~\cite[Sec. II and III]{supp} for details) and the signal PSD from an MLI with Fabry-P\'erot arm cavities (see~\cite[Sec. IV]{supp} for details) in the presence of SFs. 
 
We thus rewrite the PSD in terms of the corresponding \textit{interferometer response function} $\widetilde{\chi}_{\textsc{i}}(f,\kvec)$ as an integral over a 3-dimensional reciprocal space (see~\cite[Sec. III]{supp} for details). 
\begin{align}
S(f) &= \int \rmd^{3} \kvec \: \Gs \, \widetilde{\rho}(2 \pi f,\kvec) \, \widetilde{\chi}_{\textsc{i}}(f,\kvec),
\label{eqn:PSDirf}
\end{align}
where $\widetilde{\rho}(2 \pi f,\kvec)$ is given by~\footnote{It is interesting to contrast the general 4-dimensional Fourier transform considered above with the plane wave expansions of \textit{`metric perturbations'} considered in contemporary investigations of stochastic gravitational wave backgrounds~\cite{romano17}. These investigations consider stochastic gravitational waves as perturbations of the metric in the transverse-traceless gauge, and they require such perturbation to be a sum of plane waves that satisfy the wave equation. In contrast, we do not expect the stationary Gaussian SFs in the non-relativistic limit, to satisfy the wave equation.}
\begin{align}
\nonumber \widetilde{\rho}(2 \pi f,\kvec) = \frac{1}{(2 \pi)^{4}} \, \int \rmd^{3} \vec{r}_{12} &\int_{-\infty}^{\infty} \rmd t_{12} \, \rho\left( c t_{12}, \vec{r}_{12} \right) \\
& e^{-i \left(2 \pi f t_{12} + \kvec\cdot\vec{r}_{12}\right)}, \label{eqn:2pt}
\end{align}
and the response function $\widetilde{\chi}_{\textsc{i}}(f,\kvec)$ of the MLI (Fig. \ref{fig:HoloSetup} (a)) is 
\begin{align}
&\widetilde{\chi}_{\textsc{i}}(f,\kvec)= \left(\frac{\Li}{2}\right)^{2} \, \left\vert C_{x}(f,\kvec) - C_{z}(f,\kvec) \right\vert^{2},\label{eqn:irf}
\end{align}
with 
\begin{align}
C_{j}(f,\kvec)=&e^{i f T_{+}^{(j)}} \left\{ \text{Sinc}\left(f T_{+}^{(j)}\right) + e^{\frac{2 \pi i f \Li}{c}} \text{Sinc}\left(f T_{-}^{(j)} \right)  \right\},\label{eqn:Ci}\\
T_{\pm}^{(j)}(f,\kvec)&=\frac{\pi \Li}{c} \left(1 \pm \frac{c}{2 \pi f} \, \kvec \cdot\hat{e}_{j}\right), \quad (j=x,z).
\end{align}

We also find the response function corresponding to an MLI with Fabry-P\'erot arm cavities (see Fig. \ref{fig:HoloSetup} (b) for the interferometer geometry; also see~\cite[Sec. IV]{supp} for details). The signal PSD in terms of this response function is given by
\begin{align}
S(f) &= \int \rmd^{3} \kvec \: \Gs \, \widetilde{\rho}(2 \pi f,\kvec) \, \widetilde{\chi}_{\textsc{i}}(f,\kvec) \, \widetilde{\chi}_{\textsc{fp}}(f,\kvec), 
\label{eqn:PSDligo}
\end{align}
where the Fabry-P\'erot cavity response is given by
\begin{align}
\nonumber \widetilde{\chi}_{\textsc{fp}}(f,\kvec)= &T_{\textsc{m}}^{4} \left( \frac{1}{1- \sqrt{R_{\textsc{m}}}} \right)^{4} \\ 
&\left( \frac{1}{1+ R_{\textsc{m}}- 2 \sqrt{R_{\textsc{m}}} \:\cos (2 \pi  f/ \fsr)} \right). \label{eqn:RFcav}
\end{align}

To properly assess and compare the behaviour of the computed signal PSDs for the different correlation function classes, we consider the dimensionless frequency and PSDs,
\begin{align}
\nu&\defeq\pi f/(2 \fsr) \label{eqn:scaleFreq} \\
S_{\textsc{nc}}(\nu)&\defeq\,\left(\frac{c}{\Gs \, \Li^{3}}\right) \, S(f), \text{ and} \label{eqn:scaleA}\\
S_{\textsc{c}}(\nu)&\defeq\, \left(\frac{c}{\Gs \, \ell_{r} \, \Li^{2}} \right) \, S(f).  \label{eqn:scaleB}
\end{align}
Note that we have two types of dimensionless PSDs: $S_{\textsc{nc}}$ independent of $\ell_{r}$, catering to correlation class (a) which is also independent of $\ell_{r}$, and $S_{\textsc{c}}$ which depends on $\ell_{r}$ for correlation classes (b) and (c). 
From Eqs. \eref{eqn:PSDholo}, \eref{eqn:PSDirf}, and \eref{eqn:PSDligo}, it is clear that both $S_{\textsc{nc}}(\nu)$ and $S_{\textsc{c}}(\nu)$ are independent of $\Gs$.

\section{Results}
We present the output signal PSDs for an MLI without (see Fig. \ref{fig:Corr} (a)-(c)) and with arm cavities (see Fig. \ref{fig:HoloLIGOcomp}) for $\rho$ corresponding to correlation classes: (a) factorised $\rho_{\textsc{f}}$, (b) inverse $\rho_{\textsc{i}m}$, and (c) exponential $\rho_{\textsc{e}m}$ ($m=\textsc{s,st}$).

\subsection{Distinguishing correlation functions}

\begin{figure}
\begin{minipage}{0.48\textwidth}
\begin{minipage}{0.95\textwidth}
\includegraphics[width=\textwidth]{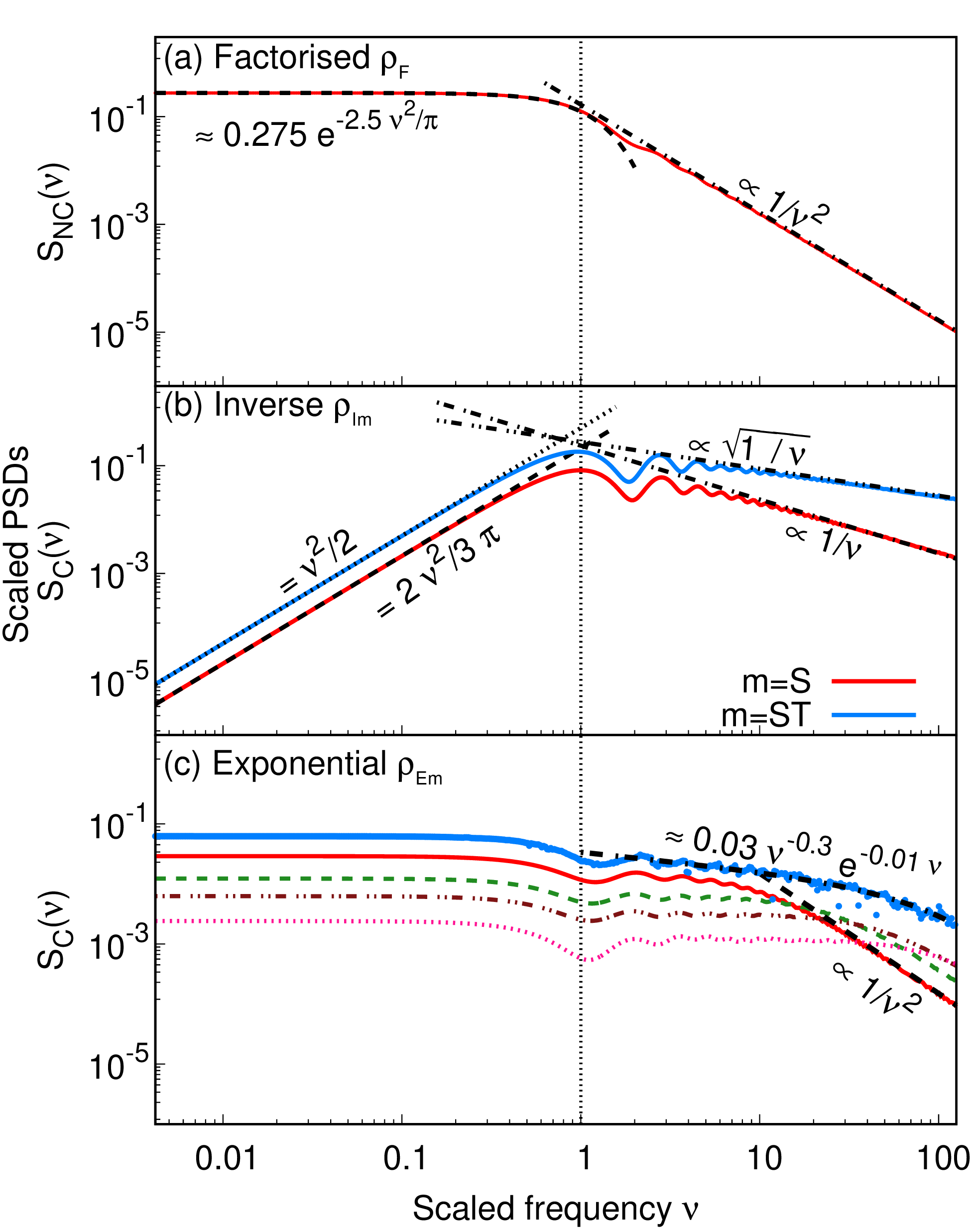}
\end{minipage}%
\hfill
\begin{minipage}{0.045\textwidth}
\rotatebox{90}{\noindent \hspace*{-0.05 em} $\xleftarrow{\phantom{MMx}}$ $\rat$ dependent $\xrightarrow{\phantom{MMx}}$ \hspace*{-0.5 em} $\xleftarrow{\phantom{MMMMMM}}$ $\Li$ independent $\xrightarrow{\phantom{MMMMMx}}$}
\end{minipage}
\end{minipage}
\caption{Output signal power spectral densities (PSDs) of an MLI due to SFs. For correlation class 
(a), \textit{factorised} $\rho_{\textsc{f}}$, the scaled PSD $S_{\textsc{nc}}(\nu)=\frac{c S(f)}{\Gs \Li^{3}} $ vs. $\nu= \frac{\pi f}{2 \fsr}$ is plotted.
For classes (b), \textit{inverse} $\rho_{\textsc{i}m},$ and 
(c) \textit{exponential} $\rho_{\textsc{e}m}$ with $m=\textsc{s,st}$, the scaled PSD $S_{\textsc{c}}(\nu)=\frac{c S(f)}{\Gs \ell_{r} \Li^{2}}$ vs $\nu$ is plotted. 
Here $m=\textsc{s}$ ($m=\textsc{st}$) denotes correlations depending on spatial (spacetime) separation. 
In (c), PSDs corresponding to $\rhoes$ for $\rat=\ell_{r}/\Li=0.025$ (red solid), $\rat=0.01$ (green dashed), $\rat=0.005$ (brown dot-dot-dashed), and $\rat=0.0025$ (pink dotted), demonstrate dependence on $\rat$.  
The PSD corresponding to $\rhoest$ for $\rat=0.025$ (blue points) is also plotted in (c). Small and large $\nu$ trends are as indicated by black dashed/dotted lines in (a)-(c). The black vertical line marks $\nu=1$.}
\label{fig:Corr}
\end{figure}

To list the three charactersitic signatures of the interferometric output signal PSD corresponding to each correlation class, we first consider the case of a simple MLI without arm cavities as in Fig. \ref{fig:HoloSetup} (a). We present the analytical expressions for the PSDs corresponding to each class in~\cite[Sec. V]{supp} (for class (a)) and \cite[Sec. VI]{supp} (for classes (b) and (c)).

\textit{Low-frequency limit $\nu\ll1$:}  For class (b) (Fig. \ref{fig:Corr} (b)), the PSDs are proportional to $\nu^{2}$ with $S_{\textsc{c}}(\nu=0)=0$ in this limit. We analytically find the constants of proportionality as $2/(3 \pi)$ and $1/2$ for $\rhois$ and $\rhoist$ respectively (see, \cite[Sec. VI]{supp} for details). This can also be understood intuitively, especially in the case of $\rhois$. In this particular case, the fluctuations satisfy the wave equation and therefore, the 4-d Fourier transform (Eq. \eref{eqn:2pt}) of $\rhois$ involves a Dirac delta function that forces $2 \pi f = c|\kvec|$ (see \cite[Eq. (81)]{supp} for the exact form). 
This leaves the interferometer response function (Eq. \eref{eqn:irf}) as the only frequency-dependent contribution to the PSD, which, up to a constant factor, is $\Li^{2} f^{2}|T_{+}^{(x)}-T_{+}^{(z)}|^{2}/4$ in this limit. The $f^{2}$ in this term produces the quadratic trend of PSD in class (b).  

For classes (a) and (c), the PSDs are almost flat in a log-log plot in the limit $\nu\ll1$ in Fig. \ref{fig:Corr} (a) and (c). We find the numerical fit in this limit for both cases to be of the form $e^{-\alpha \nu^{2}}$ where $\alpha\in\mathbb{R}$ is found numerically. In class (a), the exact numerical fit is $S_{\textsc{nc}}(\nu\ll1)\approx 0.275 e^{- \frac{5 \nu^{2}}{2 \pi}}$. This fit is independent of the value of $\Li$ (see \cite[Sec. V]{supp} for details). In class (c), the value $\alpha$ in $e^{-\alpha \nu^{2}}$ depends on the specific value of the ratio $\rat=\ell_{r}/\Li$. For instance, $\alpha=4/\pi$ for $\rat=0.01$ (see \cite[Sec. VI]{supp} for details).

\begin{figure}
\includegraphics[width=0.45\textwidth]{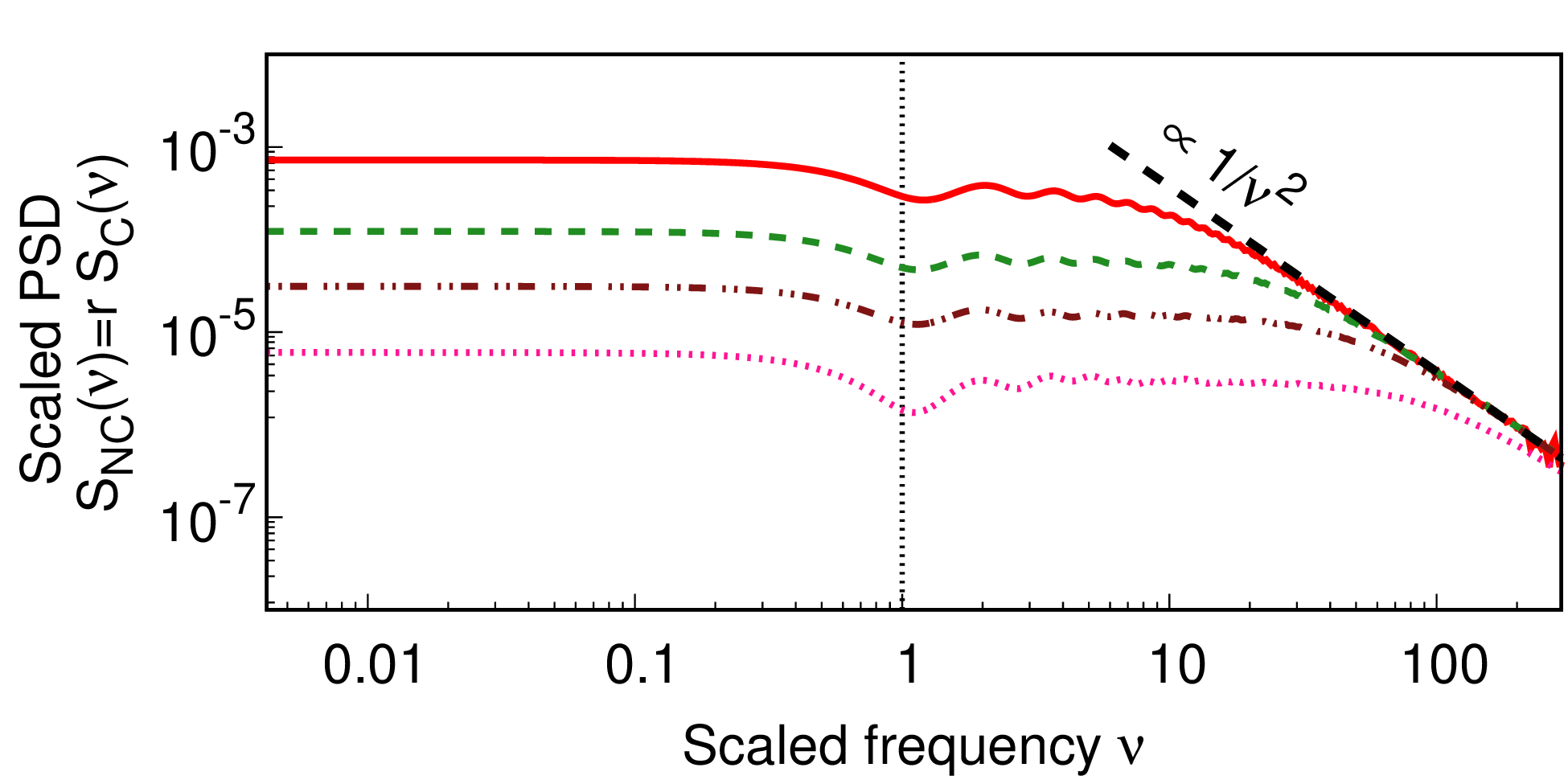}
\caption{Scaled output signal PSDs $S_{\textsc{nc}}(\nu)=\rat S_{\textsc{c}}(\nu)$ vs $\nu$ are plotted for \textit{exponential} $\rhoes$ corresponding to $\rat=0.025$ (red solid), $\rat=0.01$ (green dashed), $\rat=0.005$ (brown dot-dot-dashed), and $\rat=0.0025$ (pink dotted), demonstrating the $1/\nu^{2}$ behaviour at the high-frequency limit.}
\label{fig:HoloRhoesScale}
\end{figure}

\textit{High-frequency limit $\nu\gg 1$:}   As is evident from Fig. \ref{fig:Corr} (a)-(c), the PSDs for all correlation classes decrease with increasing $\nu$. However, the rates at which they decay are starkly different. In~\cite[Sec. VI]{supp}, we analytically justify why we find the following decay rates of $S_{\textsc{c}}(\nu)$: $\propto 1/\nu$ for $\rhois$, $\propto 1/\sqrt{\nu}$ for $\rhoist$, and $\propto 1/\nu^{2}$ for $\rhoes$. We also present $S_{\textsc{nc}}(\nu)$ vs $\nu$ for $\rhoes$ in Fig. \ref{fig:HoloRhoesScale}. Though the PSD in this case is evidently dependent on $\rat$, and therefore $\ell_{r}$, we use $S_{\textsc{nc}}(\nu)$ to showcase the high-frequency behaviour which is always $\propto 1/\nu^{2}$ irrespective of the value of $\rat$ (as this is not very apparent from Fig. \ref{fig:Corr} (c)). We also note here that the onset of this decay is delayed as the ratio $\rat$ decreases. 

The other decay rates, such as $S_{\textsc{nc}}\propto 1/\nu^{2}$ for class (a) and $S_{\textsc{c}}(\nu) \approx 0.03 \nu^{-0.3} e^{-0.01 \nu}$ for $\rhoest$ are obtained through numerical fits. Of these, the functional form of the latter is obtained by considering the possible form of the cosine transform of $\rhoest$ (see~\cite[Sec. VI]{supp} for details). 

\textit{Dependence on $\Li$ and} $\rat${\it :} For class (a), we analytically show that $S_{\textsc{nc}}(\nu)$ is independent of $\Li$ in~\cite[Eq. (68)]{supp}, and
 that $S_{\textsc{c}}(\nu)$ for class (b) is independent of $\Li$ in~\cite[Eqs. (69) and (70)]{supp}. For class (c), we analytically show that $S_{\textsc{c}}(\nu)$ depends on $\rat$ in \cite[Eqs. (74)-(76)]{supp}. Figure \ref{fig:Corr} (c) illustrates this $\rat$-dependence of the PSDs. 

As shown above, the different classes of spacetime fluctuations produce three characteristic signatures in their corresponding output signal PSDs. These can be used to identify the nature of the underlying SFs from interferometric data. To observe all three signatures characteristic to each correlation class, the interferometer should ideally be sensitive in the range of $\nu\approx0.1$ to $\nu\approx 10$. In QUEST with $\Li=3$ m (resp., GQuEST with $\Li=5$ m ), the sensitive bandwidth is designed to span from 1 MHz to 250 MHz~\cite{Sander2021} (resp., 8 MHz to 40 MHz~\cite{smv24}), with a corresponding span of $0.03\leqslant \nu \leqslant 78$ (resp., $0.42\leqslant \nu \leqslant 2.1$). This illustrates that the bandwidths of both QUEST and GQuEST would allow observation of all three signatures, although the narrower bandwidth of GQuEST could limit the observation of the low- and high-frequency signatures to some extent.

On the other hand, experimental data from LIGO covers the frequency range only from about 10 Hz to 10 kHz, corresponding roughly to $0.0004<\nu<0.4$. However, for completeness, we discuss the low- and high-frequency behaviour of the PSD for LIGO in \cite[Sec. VIII]{supp}.

To summarise, QUEST and GQuEST with their broader bandwidths allow observation of all the characteristic signatures that could help in distinguishing between correlation functions using their interferometric PSD data.

\subsection{Detecting SFs}

\begin{figure}
\includegraphics[width=0.45\textwidth]{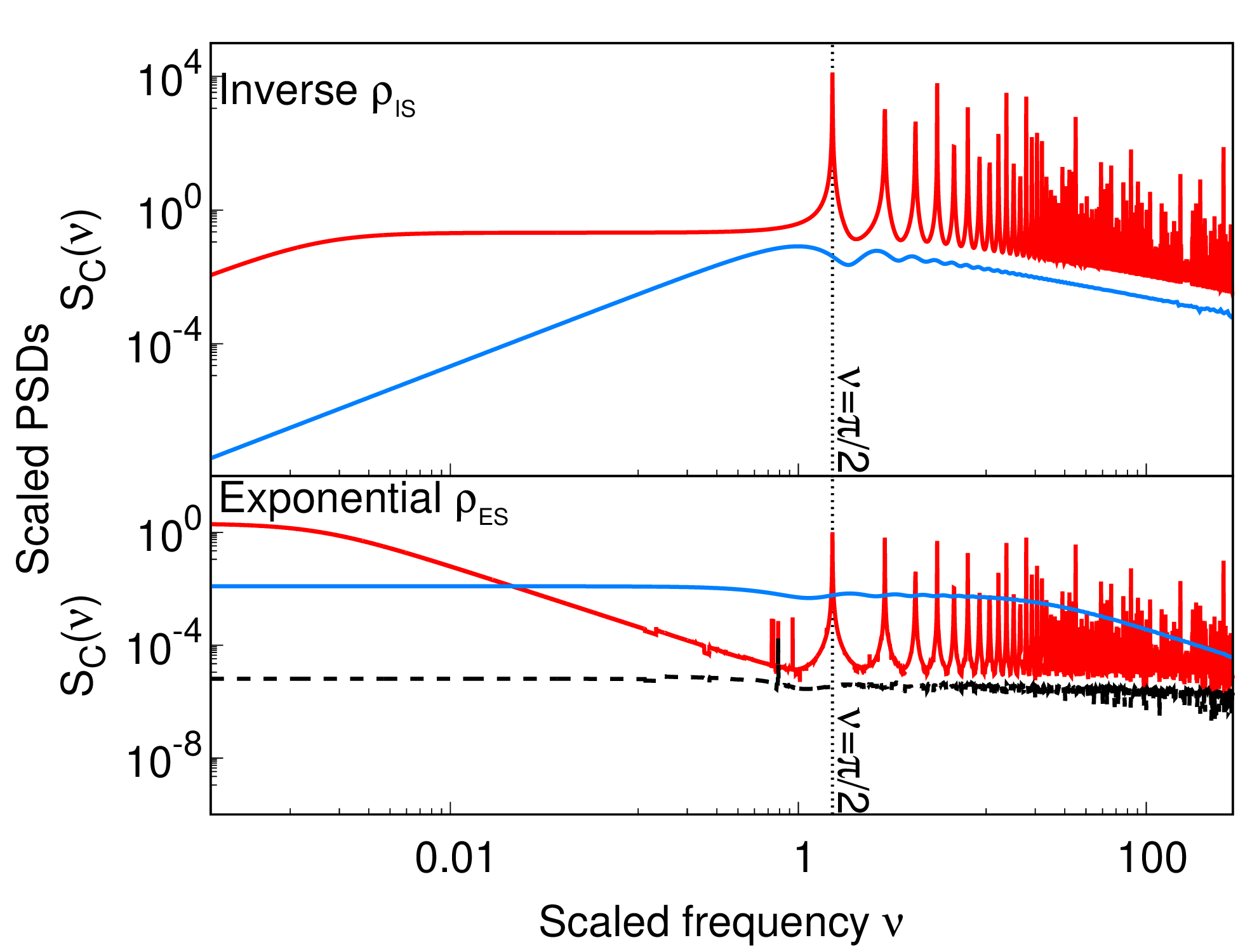}
\caption{Scaled PSD $S_{\textsc{c}}(\nu)$ corresponding to LIGO (red) with $\Li=4000$ m, QUEST (blue) with $\Li=3$ m and a hypothetical MLI without arm cavities (black dashed) with $\Li=4000$ m vs scaled frequency $\nu$ is plotted for (top) \textit{inverse} $\rhois$, and (bottom) \textit{exponential} $\rhoes$ correlation functions with $\ell_{r}=0.03$ m.}
\label{fig:HoloLIGOcomp}
\end{figure}

To highlight the advantage LIGO enjoys in detecting the SFs, we list the key features of the interferometric output signal of the MLIs with arm cavities (see Fig.~\ref{fig:HoloSetup} (b) for the geometry, and see Fig.~\ref{fig:HoloLIGOcomp} for the PSDs) for the different correlation classes. For further details, see~\cite[Sec. VI]{supp}.
\begin{enumerate}
\item \label{itm:peakgain} For any fluctuation described by a Gaussian random process, the signal PSD (Eq. \eref{eqn:PSDligo}) of MLIs with arm cavities has peaks at $\nu= m \pi/2$, for $m=1,2,3,\cdots$, of magnitude $T_{\textsc{m}}^{4} /\left( 1- \sqrt{R_{\textsc{m}}}\right)^{6}$.
For LIGO with $T_{\textsc{m}}=1-R_{\textsc{m}}=0.014$ of the input mirror of the arm cavity, the magnitude of the peak is $\approx 3.2\times 10^{5}$.
\item For class (b), $S_{\textsc{c}}$ for an MLI without arm cavities is independent of $\Li$ with a global maximum at $\nu=1$. Therefore, in this class, the Fabry-P\'erot cavity response (Eq.~\eref{eqn:RFcav}) enhances the signal strenth (Eq.~\eref{eqn:PSDligo}) at every frequency $\nu$. The strongest signal is at $\nu=\pi/2$ (equivalently, $f\approx 37.5$ kHz for LIGO), as expected. This is illustrated in the top panel of Fig. \ref{fig:HoloLIGOcomp}. This is also consistent with a prior work~\cite{geontropy2023}.
\item \label{itm:classC} For class (c),  $S_{\textsc{c}}$ is directly proportional to $\rat$. We compare $S_{\textsc{c}}$ for a given $\ell_{r}$ in two different setups, with and without arm cavities and with different arm lengths: LIGO with $\Li=4$ km, and QUEST with $\Li=3$ m. It is evident from the values of $\Li$ that the ratio $\rat$ in QUEST is far greater than that in LIGO for a given $\ell_{r}$. Thus, $S_{\textsc{c}}\propto \rat$ implies that the LIGO signal is reduced with respect to that of QUEST by the factor $\approx 10^{-3}$ (i.e., the ratio of $\Li$ of QUEST to that of LIGO). This should be considered in conjunction with Feature \ref{itm:peakgain} in this list (the presence of a peak at $\nu=\pi/2$ in the LIGO signal is always enhanced by the Fabry-P\'erot cavity gain $\approx 10^{5}$). Therefore, when measuring SF with any $\ell_{r}$ in these setups, $S_{\textsc{c}}$ at $\nu=\pi/2$ of LIGO has a peak that exceeds the $S_{\textsc{c}}$ of QUEST. This is illustrated in the bottom panel of Fig. \ref{fig:HoloLIGOcomp}.
\end{enumerate}

It is thus evident that for the classes (a)-(c) considered, LIGO has a clear advantage over QUEST and GQuEST in detecting the presence of SFs. 

It is also evident from our arguments that this advantage of LIGO is not guaranteed for all correlation functions. For instance, let us assume some class of correlation function for which $S_{\textsc{c}}$ is proportional $\rat^{2}$ in the case of an MLI without arm cavities. For such a class of correlation functions, using arguments similar to those used in discussing Feature \ref{itm:classC}, we can see that the peak in the LIGO signal does not exceed the PSD of QUEST. Therefore, LIGO does not enjoy an advantage in detection for such a class. This line of thinking might help in understanding some gravity models that predict LIGO should not enjoy any advantage in detecting SFs~\cite{kwon2016,VerZurek21}.

\section{Summary}

We have developed a methodology to systematically compute interferometric output signal power spectral densities, produced by statistically defined spacetime fluctuations (SFs), under an explicit list of assumptions. 
Using this methodology, we have compiled the interferometric output signals due to SFs in three correlation classes, for Michelson laser interferometers with and without arm cavities. This allows us to identify characteristic signatures in spectral densities for the different classes of correlation functions of SFs. We also find that (1) the laboratory-scale QUEST and GQuEST will have the broad bandwidth needed to observe all the characteristic signatures, while (2) LIGO is better suited for detecting the bare presence or absence of SFs. 

Moreover, our methodology enables unambiguous computation of interferometric signals for other (current or future) theories of gravity just from the correlation function of the SFs and the geometry of the interferometer. 
It can also be applied to compute interferometric signals to search for stochastic gravitational waves~\cite{stochGW1,stochGW2} or dark matter~\cite{LeeZurek2024}. Lastly, our methodology may be applied in instrumental `noise hunting' or calibration efforts for interferometers, for cases where the noise or calibration signal can be described as metric or phase fluctuations~\cite{GasNoise2,whitcomb_1984,weiss_ligo-t2200336} along the light path.

\section*{Acknowledgments}
We thank Jonathan Oppenheim and Ohkyung Kwon for extensive discussions and suggestions crucial to this work. 
We also thank Vincent Lee for clarifications on the Pixellon model.
BS thanks Dr. V . Balakrishnan for vital clarifications and discussions on aspects of stationarity.
BS and AD acknowledge the UK STFC ``Quantum Technologies for Fundamental Physics" program 
(Grant Numbers ST/T006404/1, ST/W006308/1 and ST/Y004493/1) for support. 
BS also acknowledges the support of the Leverhulme Trust under research grant ECF-2024-124.
SMV acknowledges the support of the Leverhulme Trust under research grant RPG-2019-022.


%
%

\bibliography{references}

\begin{thebibliography}{48}
\bibitem [{LIG()}]{LIGOunpub}%
  \BibitemOpen
  \href@noop {} {}\bibinfo {note} {{GWTC-3} run \cite{aLIGOdata} in {LIGO}
  reports data only up to 5 {kHz}. We await new results from experimental runs
  that report at higher frequency range.}\BibitemShut {Stop}%
\end{thebibliography}

\end{document}


\title{Supplementary material for ``Signatures of Correlation of Spacetime Fluctuations in Laser Interferometers"}

\author{B.~Sharmila}
\email{Sharmila.Balamurugan@warwick.ac.uk}
\affiliation{Department of Physics, University of Warwick, Coventry CV4 7AL, UK.}

\author{Sander M. Vermeulen}
\email{smv@caltech.edu}
\affiliation{Division of Physics, Mathematics and Astronomy,
California Institute of Technology, Pasadena, CA 91125, USA.}

\author{Animesh Datta}
\email{Animesh.Datta@warwick.ac.uk}
\affiliation{Department of Physics, University of Warwick, Coventry CV4 7AL, UK.}

\date{\today}

\maketitle


\section{I. Propagation of light in a fluctuating spacetime}

In this section, we will solve the relativistic wave equation, applying necessary approximations, to obtain the electric field of light propagating in a fluctuating spacetime.

The relativistic wave equation in terms of the electromagnetic field tensor $F_{\alpha \beta}$~\cite{Tsagas2005}
\begin{equation}
\nonumber \Box F_{\alpha \beta} + 2 R_{\alpha \gamma \beta \delta} F^{\gamma \delta} - R_{\alpha \gamma} F^{\gamma}_{\beta} + R_{\beta \gamma} F^{\gamma}_{\alpha} =0,
\label{eqn:relativWE}
\end{equation}
where
\begin{eqnarray}
&&\nonumber \Box F_{\alpha \beta} = g^{\gamma \delta} \nabla_{\gamma} \nabla_{\delta} F_{\alpha \beta} \\
&&\nonumber =g^{\gamma \delta} \bigg( \partial_{\delta} X_{\alpha \beta \gamma} - \Gamma^{\eta}_{\alpha \delta} X_{\eta \beta \gamma} - \Gamma^{\eta}_{\beta \delta} X_{\alpha \eta \gamma}\\
&& \hspace*{10 em} - \Gamma^{\eta}_{\gamma\delta} X_{\alpha \beta \eta} \bigg),
\end{eqnarray}
with
\be
X_{\alpha \beta \gamma}=\partial_{\gamma} F_{\alpha \beta} - \Gamma^{\eta}_{\beta \gamma} F_{\alpha \eta} - \Gamma^{\eta}_{\alpha \gamma} F_{\eta \beta}, 
\ee
\be
R_{\alpha \beta \gamma \delta} = g_{\alpha \nu} \left[ \partial_{\gamma} \Gamma^{\nu}_{\delta \beta} - \partial_{\delta} \Gamma^{\nu}_{\gamma \beta} + \Gamma^{\nu}_{\gamma \eta} \Gamma^{\eta}_{\delta \beta} - \Gamma^{\nu}_{\delta \eta} \Gamma^{\eta}_{\gamma \beta}\right],
\ee
and
\be
R_{\alpha \beta}= R^{\nu}_{\alpha \nu \beta},
\ee
with the Christoffel symbol 
\be
\Gamma^{\alpha}_{\beta \gamma}=g^{\alpha \mu} (\partial_{\beta} g_{\mu \gamma}+\partial_{\gamma} g_{\beta \mu}-\partial_{\mu} g_{\beta \gamma})/2.
\ee
Here the Greek indices take values from the set $\{0,1,2,3\}$, with 0 corresponding to the timelike component and the rest to spacelike components.

Assuming the most general $g_{\beta \gamma}$ without any further assumptions, it can be trivially seen that it is not possible to simplify the above relativistic equation. Therefore, we consider the eikonal approximation to find a solution to $-c F_{0j}$ ($j=1,2,3$) or equivalently, $\vec{E}$, the 3-vector electric field. 
\begin{enumerate}[label=(\roman*)]
\item \label{Ass:eikonal} \textit{Ansatz:} Let us consider electric field of the form,
\begin{eqnarray}
\vec{E}(\bm{r})=\vec{E}_{0}(\bm{r}) e^{i k \Phi(\bm{r})},
\label{eqn:Eans}
\end{eqnarray}
where $k=2 \pi/\lambda=\Omega/c$ with the wavelength $\lambda$ and the frequency $\Omega$ of the electromagnetic (EM) radiation propagating in the fluctuating spacetime. Also $\bm{r}\equiv(t,x,y,z)$.
To apply the eikonal approximation, we use Eq. \eref{eqn:Eans} in Eq. \eref{eqn:relativWE} and consider $k\to\infty$. Note that this sets the wavelength and the time-period of the EM radiation to be the smallest length and time scales respectively in the system. 
\end{enumerate}

Using Assumption \ref{Ass:eikonal}, we find that Eq. \eref{eqn:relativWE} reduces to
\begin{equation}
g^{\gamma \delta} \partial_{\gamma} \partial_{\delta} F_{\alpha \beta}=0.
\end{equation}
This is because in the presence of terms that are 2-order derivatives of $F_{\alpha \beta}$, terms proportional to smaller order derivatives don't survive.

Let us now consider the following form of the metric to simplify this further. 
\begin{enumerate}[label=(\arabic*)]
\item \label{mod:metric} We consider a spacetime metric of the form,
\begin{eqnarray}
g^{\mu \nu}=\eta^{\mu \nu} + 2 w^{\mu \nu},
\end{eqnarray}
where the $4\times 4$ matrix $w$ is a real, symmetric, matrix that models fluctuations about the flat Minkowski metric $\eta_{\mu \nu}$ with a signature $(-1,1,1,1)$.
\end{enumerate}

Applying both Assumption \ref{Ass:eikonal} and Attribute \ref{mod:metric}, we find
\begin{align}
&\nonumber -\frac{(1-2 w^{00})}{c^{2}} (\partial_{t} \Phi)^{2} + (1+2 w^{11}) (\partial_{x} \Phi)^{2}\\
&\nonumber + (1+2 w^{22}) (\partial_{y} \Phi)^{2} + (1+2 w^{33}) (\partial_{z} \Phi)^{2}\\ 
&\nonumber + \frac{4}{c} (\partial_{t} \Phi) \left[ w^{01} (\partial_{x} \Phi) + w^{02} (\partial_{y} \Phi) + w^{03} (\partial_{z} \Phi) \right]\\
&+ 4 (\partial_{x} \Phi) \left[ w^{12} (\partial_{y} \Phi) + w^{13} (\partial_{z} \Phi) \right] + 4 w^{23} (\partial_{y} \Phi) (\partial_{z} \Phi) = 0.
\end{align}

\begin{enumerate}[label=(\roman*)]
\setcounter{enumi}{1}
\item \label{Ass:PhiAns} \textit{Slowly varying envelope approximation (SVEA):} To be consistent with Assumption \ref{Ass:eikonal}, we also consider the following ansatz,
\begin{eqnarray}
\Phi(\bm{r})=c t - z + \Phi_{\textsc{f}}(\bm{r}),
\label{eqn:PhiAns}
\end{eqnarray}
with $\partial_{\mu} \Phi_{\textsc{f}} \ll 1$ ($\mu=0,1,2,3$).
\end{enumerate}
Using Assumption \ref{Ass:PhiAns}, we neglect terms of order $(\partial_{\mu} \Phi_{\textsc{f}})^{2}$. With this, we find
\begin{align}
&\nonumber -(1-2 w^{00}) -\frac{2 (1-2 w^{00})}{c} (\partial_{t} \Phi_{\textsc{f}}) \\
&\nonumber + (1+2 w^{33}) - 2 (1+2 w^{33}) (\partial_{z} \Phi_{\textsc{f}}) + 4 w^{01} (\partial_{x} \Phi_{\textsc{f}})\\ 
&\nonumber + 4 w^{02} (\partial_{y} \Phi_{\textsc{f}}) + 4 w^{03} (\partial_{z} \Phi_{\textsc{f}}) - \frac{4}{c} w^{03} (\partial_{t} \Phi_{\textsc{f}}) - 4 w^{03} \\
&- 4 w^{13} (\partial_{x} \Phi_{\textsc{f}}) - 4 w^{23} (\partial_{y} \Phi_{\textsc{f}}) = 0.
\end{align}

\begin{enumerate}[label=(\arabic*)]
\setcounter{enumi}{1}
\item \label{mod:smallFluct} We also consider the metric fluctuations to be small, i.e., $w^{\mu \nu} \ll 1$ ($\mu,\nu=0,1,2,3$).
\end{enumerate}
Using Assumption \ref{Ass:PhiAns} and Attribute \ref{mod:smallFluct} together, we neglect terms of order $w^{\mu \nu} (\partial_{\gamma} \Phi_{\textsc{f}})$. With this, we find
\begin{align}
-\frac{2}{c} (\partial_{t} \Phi_{\textsc{f}}) - 2 (\partial_{z} \Phi_{\textsc{f}}) + 2 w^{00} + 2 w^{33} - 4 w^{03} = 0.
\end{align}

A general solution is
\begin{align}
\nonumber \Phi_{\textsc{f}}(\bm{r})= \mathcal{F}\left(c t -z \right) + c \int\limits_{0}^{t} \text{d}t' \, \left[ w^{00}\left(\bm{r}(t')\right) \right. \\
\left. + w^{33}\left(\bm{r}(t')\right) - 2 w^{03}\left(\bm{r}(t')\right) \right],  
\end{align}
where $\mathcal{F}$ denotes any general function of the given argument.
We choose the following solution as one that best fits our initial conditions of an input Gaussian beam.
\begin{eqnarray}
\vec{E}(\bm{r})=\vec{E}_{in}(x,y) e^{i k \Phi(\bm{r})},
\label{eqn:Soln}
\end{eqnarray}
where
\begin{subequations}
\begin{align}
\nonumber &\vec{E}_{in}(x,y) = \sqrt{\frac{2}{\pi} \,} \, \frac{z_{R}}{W_{0} \sqrt{z_{R}^{2} + z_{0}^{2}}}\\
& \hspace*{3 em} \exp \left[ \left( -\frac{i k z_{0} W_{0}^{2} + 2 z_{R}^{2} }{2 W_{0}^{2} (z_{0}^{2}+z_{R}^{2})}  \right) (x^{2} + y^{2}) \right] \hat{e}_{y},\\
\nonumber &\Phi(\bm{r})=c t -z +c  \int\limits_{0}^{t} \text{d}t' \, \left[ w^{00}\left(\bm{r}(t')\right) + w^{33}\left(\bm{r}(t')\right)  \right. \\
&  \hspace*{13 em} \left. - 2 w^{03}\left(\bm{r}(t')\right) \right],  
\end{align}
\label{eqn:APhi}
\end{subequations}
with $\hat{e}_{y}$ being the unit vector along the $y$-axis, $W_{0}$ the beam waist, $z_{0}$ the position of the beam waist, and $z_{R}=\pi W_{0}^{2}/\lambda$. 

\section{II. Holometer-type setup: Spectral densities}

In this section, we obtain the power and cross spectral densities of the optical path difference between the two arms of the Michelson laser interferometer (MLI) with no arm cavities.

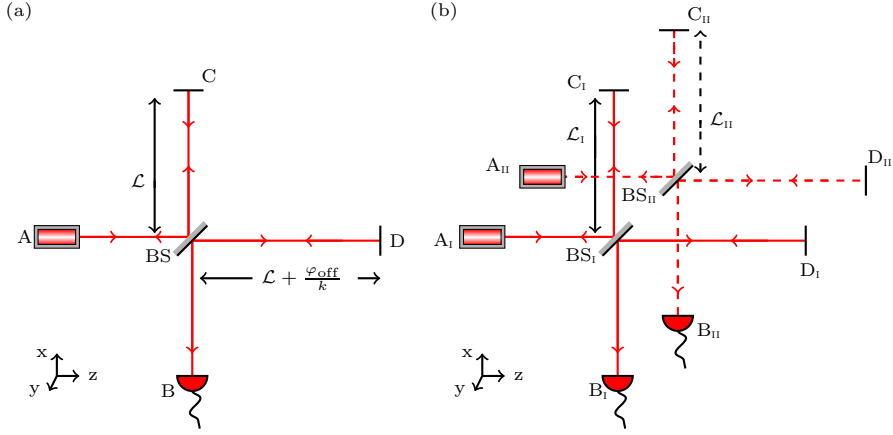
\begin{figure*}
\begin{tikzpicture}
\draw[black,thick] (0.,3.05) node[left] (scrip) {\scriptsize
(a)};
\draw[red,thick] (0.,0.05) -- (1.95,0.05);
\draw[red,thick] (2.,0.) -- (4.5,0.);
\draw[red,thick] (2.,-2.) -- (2.,0.);
\draw[red,thick] (1.95,0.05) -- (1.95,2.);
\draw[red,thick,->] (0.,0.05) -- (1.,0.05);
\draw[red,thick,->] (2.,0.) -- (3.,0.);
\draw[red,thick,->] (1.95,0.05) -- (1.5,0.05);
\draw[red,thick,->] (4.,0.) -- (3.5,0.);
\draw[red,thick,->] (2.,0.) -- (2.,-1.5);
\draw[red,thick,->] (1.95,0.05) -- (1.95,1.);
\draw[red,thick,->] (1.95,2.) -- (1.95,1.5);
\draw[black,thick] (4.5,-0.2) -- (4.5,0.2);
\draw[black,thick] (1.75,2.) -- (2.15,2.);
\draw[black,thick] (1.85,-0.2) node[left] (scrip) {\scriptsize
 BS};
\draw[black!30,fill=black!30] (1.8,-0.2) -- (2.2,0.2)--(2.15,0.25)--(1.75,-0.15)--cycle;
\draw[black,thick] (1.8,-0.2) -- (2.2,0.2);
\draw[black,thick,<-] (2.1,-0.5) -- (2.8,-0.5) node[right] (scrip) {\scriptsize
 $\Li+\frac{\varphi_{\text{off}}}{k}$};
\draw[black,thick,->] (4.2,-0.5) -- (4.5,-0.5);
\draw[black,thick,<-] (1.5,0.1) -- (1.5,0.8) node[left] (scrip) {\scriptsize
 $\Li$};
\draw[black,thick,->] (1.5,0.7) -- (1.5,1.9);
\draw[black,fill=black!30] (-0.1,-0.1) rectangle (0.5,0.2);
\draw[lasernode] (-0.05,-0.05) rectangle (0.45,0.15);
\draw[black,thick,fill=red] (2.2,-1.8) -- (1.8,-1.8) arc(180:360:0.2) --cycle;
\draw[black,thick,snake it] (2,-2) -- (2.1,-2.5);
\draw[black,thick] (0.,0.05) node[left] (scrip) {\scriptsize
A};
\draw[black,thick] (1.9,-2.) node[left] (scrip) {\scriptsize
B};
\draw[black,thick] (4.5,0.) node[right] (scrip) {\scriptsize
D};
\draw[black,thick] (2.,2.2) node[right] (scrip) {\scriptsize
C};
\draw[black,thick,->] (0.2,-1.8) -- (0.5,-1.8) node[right] (scrip) {\scriptsize
z};
\draw[black,thick,->] (0.2,-1.8) -- (0.2,-1.5) node[left] (scrip) {\scriptsize
x};
\draw[black,thick,->] (0.2,-1.8) -- (0.1,-2) node[left] (scrip) {\scriptsize
y};
\end{tikzpicture}
\begin{tikzpicture}
\draw[black,thick] (0.,3.05) node[left] (scrip) {\scriptsize
(b)};
\draw[red,thick] (0.,0.05) -- (1.95,0.05);
\draw[red,thick] (2.,0.) -- (4.5,0.);
\draw[red,thick] (2.,-2.) -- (2.,0.);
\draw[red,thick] (1.95,0.05) -- (1.95,2.);
\draw[red,thick,->] (0.,0.05) -- (1.,0.05);
\draw[red,thick,->] (2.,0.) -- (3.,0.);
\draw[red,thick,->] (1.95,0.05) -- (1.5,0.05);
\draw[red,thick,->] (4.,0.) -- (3.5,0.);
\draw[red,thick,->] (2.,0.) -- (2.,-1.5);
\draw[red,thick,->] (1.95,0.05) -- (1.95,1.);
\draw[red,thick,->] (1.95,2.) -- (1.95,1.5);
\draw[black,thick] (4.5,-0.2) -- (4.5,0.2);
\draw[black,thick] (1.75,2.) -- (2.15,2.);
\draw[black,thick] (1.85,-0.2) node[left] (scrip) {\scriptsize 
BS$_{\textsc{i}}$};
\draw[black!30,fill=black!30] (1.8,-0.2) -- (2.2,0.2)--(2.15,0.25)--(1.75,-0.15)--cycle;
\draw[black,thick] (1.8,-0.2) -- (2.2,0.2);
\draw[black,thick,<-] (1.7,0.1) -- (1.7,1.4) node[left] (scrip) {\scriptsize
 $\Li_{\textsc{i}}$};
\draw[black,thick,->] (1.7,1.1) -- (1.7,1.9);
\draw[black,fill=black!30] (-0.1,-0.1) rectangle (0.5,0.2);
\draw[lasernode] (-0.05,-0.05) rectangle (0.45,0.15);
\draw[black,thick,fill=red] (2.2,-1.8) -- (1.8,-1.8) arc(180:360:0.2) --cycle;
\draw[black,thick,snake it] (2,-2) -- (2.1,-2.5);
\draw[black,thick] (-0.05,0.) node[left] (scrip) {\scriptsize 
A$_\textsc{i}$};
\draw[black,thick] (2.,-2.) node[left] (scrip) {\scriptsize
 B$_\textsc{i}$};
\draw[black,thick] (4.3,-0.4) node[right] (scrip) {\scriptsize
 D$_\textsc{i}$};
\draw[black,thick] (1.2,2.1) node[right] (scrip) {\scriptsize
 C$_\textsc{i}$};
\draw[red,thick,dashed] (0.8,0.85) -- (2.75,0.85);
\draw[red,thick,dashed] (2.8,0.8) -- (5.3,0.8);
\draw[red,thick,dashed] (2.8,-1.2) -- (2.8,0.8);
\draw[red,thick,dashed] (2.75,0.85) -- (2.75,2.8);
\draw[red,thick,dashed,->] (0.8,0.85) -- (1.6,0.85);
\draw[red,thick,dashed,->] (2.8,0.8) -- (3.8,0.8);
\draw[red,thick,dashed,->] (2.75,0.85) -- (2.3,0.85);
\draw[red,thick,dashed,->] (4.8,0.8) -- (4.3,0.8);
\draw[red,thick,dashed,->] (2.8,0.8) -- (2.8,-0.7);
\draw[red,thick,dashed,->] (2.75,0.85) -- (2.75,1.8);
\draw[red,thick,dashed,->] (2.75,2.8) -- (2.75,2.3);
\draw[black,thick] (5.3,0.6) -- (5.3,1);
\draw[black,thick] (2.55,2.8) -- (2.95,2.8);
\draw[black,thick] (2.65,0.6) node[left] (scrip) {\scriptsize
 BS$_{\textsc{ii}}$};
\draw[black!30,fill=black!30] (2.6,0.6) -- (3,1)--(2.95,1.05)--(2.55,0.65)--cycle;
\draw[black,thick] (2.6,0.6) -- (3.,1.);
\draw[black,thick,dashed,<-] (3.1,0.9) -- (3.1,1.6) node[right] (scrip) {\scriptsize
 $\Li_{\textsc{ii}}$};
\draw[black,thick,dashed,->] (3.1,1.5) -- (3.1,2.8);
\draw[black,fill=black!30] (0.7,0.7) rectangle (1.3,1);
\draw[lasernode] (0.75,0.75) rectangle (1.25,0.95);
\draw[black,thick,fill=red] (3,-1.) -- (2.6,-1.) arc(180:360:0.2) --cycle;
\draw[black,thick,snake it] (2.8,-1.2) -- (2.9,-1.7);
\draw[black,thick,dashed] (0.7,1.) node[left] (scrip) {\scriptsize
 A$_\textsc{ii}$};
\draw[black,thick,dashed] (3.5,-1.2) node[left] (scrip) {\scriptsize
 B$_\textsc{ii}$};
\draw[black,thick,dashed] (5.2,1.1) node[right] (scrip) {\scriptsize
 D$_\textsc{ii}$};
\draw[black,thick,dashed] (2.8,3.) node[right] (scrip) {\scriptsize
 C$_\textsc{ii}$};
\draw[black,thick,->] (0.2,-1.8) -- (0.5,-1.8) node[right] (scrip) {\scriptsize
z};
\draw[black,thick,->] (0.2,-1.8) -- (0.2,-1.5) node[left] (scrip) {\scriptsize
x};
\draw[black,thick,->] (0.2,-1.8) -- (0.1,-2) node[left] (scrip) {\scriptsize
y};
\end{tikzpicture}
\caption{A schematic diagram of the interferometric setup. (a) Michelson laser interferometer (MLI) with a laser source at the input port A and a detector at the output port B with the two perpendicular arms denoted by C and D. 
The 50/50 lossless beamsplitter is denoted by BS and is taken as the origin of the reference frame in our computation. We can effectively assume the detector to be at the origin as any change suffered by the light after interference at the BS is common to output field contributions from both the arms and therefore cannot be detected. (b) Two co-located MLIs with input ports A$_{i}$ and output port B$_{i}$ with the two perpendicular arms denoted by C$_{i}$ and D$_{i}$ each with arm length $\Li^{i}$ ($i=$I,II). We consider $\Li^{\textsc{i}}=\Li^{\textsc{ii}}=\Li$. The origin is at BS$_{\textsc{i}}$.
}
\label{fig:HoloSetup}
\end{figure*}

The electric field at the output port (see Fig. \ref{fig:HoloSetup})  of the Interferometer $p$ ($p=\textsc{i,ii}$) is
\begin{eqnarray}
\nonumber E^{p}_{\mathrm{out}}(\bm{r}_{\textsc{d}}(\Delta_{\tau},\Delta))= \frac{1}{\sqrt{2}}  \bigg[ E^{(\textsc{c}_{p})}_{y}(\bm{r}_{\textsc{d}}(\Delta_{\tau},\Delta))\\
- E^{(\textsc{d}_{p})}_{y}(\bm{r}_{\textsc{d}}(\Delta_{\tau},\Delta)) e^{- 2 i \varphi_{\text{off}}} \bigg]. \hspace*{2 em}
\label{eqn:Eout}
\end{eqnarray}
Here the detector is at $\bm{r}_{\textsc{d}}(\Delta_{\tau},\Delta)\equiv (\tau_{0}+\Delta_{\tau},\Delta,\Delta,\Delta)$ with $\Delta=0$ for Interferometer I and $\Delta=\Delta_{s}$ for Interferometer II. We note that $\tau_{0}=2\Li/c$.

As a first step towards finding the spectral densities of the interferometric output, we define electric field correlation tensors of the form,
\begin{widetext}
\be
M^{m,\M}_{\bm{c};\bm{c'}} \left(\{\bm{r}\}; \{\bm{r'}\} \right)
=
E^{(c_1)}_{y}(\bm{r}_1)
E^{(c_2)}_{y}(\bm{r}_2)
\cdots 
E^{(c_m)}_{y}(\bm{r}_m)
E^{(c'_1) \ast}_{y}(\bm{r'}_1) 
E^{(c'_2) \ast}_{y}(\bm{r'}_2) 
 \cdots 
E^{(c'_{\M}) \ast}_{y}(\bm{r'}_{\M})
\label{eqn:Meas}
\ee
\end{widetext}
where  
$c_i, c'_j \in \{\text{C},\text{D}\}$ for $i=1,2,\dots,m$,  $j=1,2,\dots,\M$, with $m$ not necessarily equal to $\M$ and 
$\{\bm{r}\} \equiv \{ \bm{r}_1,\cdots \bm{r}_m\},$
$\bm{c} \equiv \{ c_1,\cdots c_m\}.$
The primed variables are denoted similarly. Here $(m,\M)$ denote the order of the electric field correlation tensor. For brevity, we define the correlation function 
$M_{\mathrm{out}}^{m,\M} \left(\{\bm{r}\}; \{\bm{r'}\} \right)$ for the output field of an MLI in line with $M^{m,\M}_{\bm{c};\bm{c'}} \left(\{\bm{r}\}; \{\bm{r'}\} \right)$,
except with $E_{\mathrm{out}}$ replacing the field components $E^{(c)}_{y}$ and $E^{(c')}_{y}$ in Eq. \eref{eqn:Meas}. 

The fourth-order correlation $M^{2,2}_{\mathrm{out}} \left(\bm{R};\bm{R}\right)$ is written explicitly in terms of $M^{2,2}_{\{\textsc{x,y}\};\{\textsc{x',y'}\}} \left( \bm{R};\bm{R} \right)$ as 
\begin{widetext}
\begin{align}
\nonumber M^{2,2}_{\mathrm{out}} \left(\bm{R};\bm{R}\right)=\frac{1}{4} &\left[ \rule{0cm}{0.7cm} \sum_{\text{X}\in \{\text{C},\text{D}\}}\sum_{\text{Y}\in \{\text{C},\text{D}\}}M^{2,2}_{\{\textsc{x,y}\};\{\textsc{x,y}\}} \left( \bm{R};\bm{R} \right) - \sum_{\text{X}\in \{\text{C},\text{D}\}} M^{2,2}_{\{\textsc{x,c}\};\{\textsc{x,d}\}} \left( \bm{R};\bm{R} \right) e^{2 i \varphi_{\text{off}}} \right. \\
\nonumber & - \sum_{\text{X}\in \{\text{C},\text{D}\}} M^{2,2}_{\{\textsc{x,d}\};\{\textsc{x,c}\}} \left( \bm{R};\bm{R} \right) e^{- 2 i \varphi_{\text{off}}} - \sum_{\text{X}\in \{\text{C},\text{D}\}} M^{2,2}_{\{\textsc{c},\textsc{x}\};\{\textsc{d},\textsc{x}\}} \left( \bm{R};\bm{R} \right) e^{ 2 i \varphi_{\text{off}}} \\
& \nonumber - \sum_{\text{X}\in \{\text{C},\text{D}\}} M^{2,2}_{\{\textsc{d},\textsc{x}\};\{\textsc{c},\textsc{x}\}} \left( \bm{R};\bm{R} \right) e^{- 2 i \varphi_{\text{off}}} + M^{2,2}_{\{\textsc{c},\textsc{c}\};\{\textsc{d},\textsc{d}\}} \left( \bm{R};\bm{R} \right) e^{4 i \varphi_{\text{off}}}\\
& \left.  + M^{2,2}_{\{\textsc{d},\textsc{d}\};\{\textsc{c},\textsc{c}\}} \left( \bm{R};\bm{R} \right) e^{- 4 i \varphi_{\text{off}}} + M^{2,2}_{\{\textsc{d},\textsc{c}\};\{\textsc{c},\textsc{d}\}} \left( \bm{R};\bm{R} \right) + M^{2,2}_{\{\textsc{c},\textsc{d}\};\{\textsc{d},\textsc{c}\}} \left( \bm{R};\bm{R} \right) \rule{0cm}{0.7cm} \right].
\label{eqn:m22DC}
\end{align}

We find $\overline{M^{2,2}_{\mathrm{out}} \left(\bm{R};\bm{R}\right)}$ with $\bm{R}=\{(\tau_{0},0,0,0), (\tau_{0}+\Delta_{\tau},\Delta,\Delta,\Delta)\}$. We discuss the salient steps involved in computing $\overline{M^{2,2}_{\mathrm{out}} \left(\bm{R};\bm{R}\right)}$ by listing the steps in computing one of the terms in this moment, such as,
\begin{eqnarray}
\overline{M^{2,2}_{\{\textsc{d},\textsc{d}\};\{\textsc{c},\textsc{c}\}} \left( \bm{R};\bm{R} \right)}&=& |\vec{E}_{in}(0,0)|^{2} |\vec{E}_{in}(\Delta,\Delta)|^{2} \: \overline{\left(\substack{\exp\left[-i \Omega \left\{ \int\limits_{0}^{\tau_{0}} \text{d}t' \, w\left(t',0,0,s(t')\right) + \int\limits_{\Delta_{\tau}}^{\tau_{0}+\Delta_{\tau}} \text{d}t' \, w\left(t',\Delta,\Delta,s(t'-\Delta_{\tau})+\Delta\right) \right\}\right]\\ \exp\left[i \Omega \left\{ \int\limits_{0}^{\tau_{0}} \text{d}t' \, w\left(t',s(t'),0,0\right) + \int\limits_{\Delta_{\tau}}^{\tau_{0}+\Delta_{\tau}} \text{d}t' \, w\left(t',\Delta+s(t'-\Delta_{\tau}),\Delta,\Delta\right) \right\}\right]}\right)}.
\label{eqn:m22ddcc}
\end{eqnarray}
Here $s(t)=c t$ if $0\leqslant t \leqslant \tau_{0}/2$ and $s(t)=2 \Li - c t$ if $\tau_{0}/2<t\leqslant\tau_{0}$. We define this function up to $t\leqslant\tau_{0}$ when we consider the Holometer. We extend the definition when we consider LIGO.\\

To simplify $\overline{M^{2,2}_{\mathrm{out}} \left(\bm{R};\bm{R}\right)}$, we need to define two correlation integrals for which the following assumptions are required.
\begin{enumerate}[label=(\roman*)]
\setcounter{enumi}{2}
\item \label{Ass:stat} \textit{Stationarity assumption:} $w(\bm{r})$ is a stationary Gaussian random process with
\begin{align}
&\overline{w}=0, \quad \text{and} \\ 
&\overline{w(t_{1},\vec{r}_{1}) w(t_{2},\vec{r}_{2})}  = \Gs \, \rho\left( c t_{12} , \vec{r}_{12} \right).
\end{align}
Here $o_{12}=o_{1}-o_{2}$ ($o=t,\vec{r}$) with $\vec{r}_{i}\equiv(x_{i},y_{i},z_{i})$ ($i=1,2$) and $\Gs$ is the strength of the fluctuations.
\item \label{Ass:iso} \textit{Isotropy:} The two-point correlation function $\rho$ is isotropic in space, i.e., $\rho\left(\delta_{1}, \{ \delta_{2}, \delta_{3}, \delta_{4}\}\right) = \rho\left(\delta_{1}, \{ \delta_{4}, \delta_{2}, \delta_{3} \}\right)= \rho\left(\delta_{1}, \{ \delta_{4}, \delta_{3}, \delta_{2}\} \right) = \dots$, for any separation $\delta_{i}$ ($i=1,2,3,4$). To achieve this isotropy, we consider the correlation to decay with a correlation scale $\ell_{r}$ in all three spatial dimensions. We also additionally consider the temporal correlation scale to be $\ell_{r}/c$.  
\end{enumerate}
We define the two correlation integrals as follows.
\begin{subequations}
\begin{eqnarray}
\nonumber &&\zeta_{1}(\Delta_{\tau},\Delta)= \int\limits_{0}^{\tau_{0}} \rmd t_{1} \int\limits_{0}^{\tau_{0}} \rmd t_{2} \, \rho(t_{1}+\Delta_{\tau}-t_{2},\Delta,\Delta,s(t_{1})+\Delta-s(t_{2}))\\
&& = \int\limits_{0}^{\tau_{0}} \rmd t_{1} \int\limits_{0}^{\tau_{0}} \rmd t_{2} \, \rho(t_{1}+\Delta_{\tau}-t_{2},s(t_{1})+\Delta-s(t_{2}),\Delta,\Delta),
\label{eqn:Zeta1Defn}\\
&&\zeta_{2}(\Delta_{\tau},\Delta) = \int\limits_{0}^{\tau_{0}} \rmd t_{1} \, \int\limits_{0}^{\tau_{0}} \rmd t_{2} \, \rho(t_{1}+\Delta_{\tau}-t_{2},s(t_{1})+\Delta,\Delta,\Delta-s(t_{2})).\label{eqn:Zeta2Defn}
\end{eqnarray}
\label{eqn:CorrSet}
\end{subequations}
Here the $\sigma(\Delta_{\tau})=\zeta_{1}(\Delta_{\tau},0)$ and $\xi(\Delta_{\tau})=\zeta_{2}(\Delta_{\tau},0)$. 

Using Eqs. \eref{eqn:m22ddcc} and \eref{eqn:CorrSet}, we find
\begin{eqnarray}
\nonumber &&\overline{M^{2,2}_{\{\textsc{c},\textsc{c}\};\{\textsc{d},\textsc{d}\}} \left( \bm{R};\bm{R} \right)} =\overline{M^{2,2}_{\{\textsc{d},\textsc{d}\};\{\textsc{c},\textsc{c}\}} \left( \bm{R};\bm{R} \right)} = |\vec{E}_{in}(0,0)|^{2} |\vec{E}_{in}(\Delta,\Delta)|^{2} \: e^{- \Omega^{2} \Gs \left\{ 2 \zeta_{1}(0,\Delta)  + 2 \zeta_{1}(\Delta_{\tau},\Delta) - 2 \zeta_{2}(0,\Delta)  - 2 \zeta_{2}(\Delta_{\tau},\Delta) \right\}} \\
&&\approx |\vec{E}_{in}(0,0)|^{2} |\vec{E}_{in}(\Delta,\Delta)|^{2} \Bigg[ 1 - \Omega^{2} \Gs \bigg\{ 2 \zeta_{1}(0,\Delta)  + 2 \zeta_{1}(\Delta_{\tau},\Delta) - 2 \zeta_{2}(0,\Delta)  - 2 \zeta_{2}(\Delta_{\tau},\Delta) \bigg\}\Bigg],\\
\nonumber && \overline{M^{2,2}_{\{\textsc{c},\textsc{d}\};\{\textsc{c},\textsc{d}\}} \left( \bm{R};\bm{R} \right)} =\overline{M^{2,2}_{\{\textsc{d},\textsc{c}\};\{\textsc{d},\textsc{c}\}} \left( \bm{R};\bm{R} \right)} =|\vec{E}_{in}(0,0)|^{2} |\vec{E}_{in}(\Delta,\Delta)|^{2} \: e^{ - \Omega^{2} \Gs \left\{ 2 \zeta_{1}(0,\Delta)  - 2 \zeta_{1}(\Delta_{\tau},\Delta) - 2 \zeta_{2}(0,\Delta)  + 2 \zeta_{2}(\Delta_{\tau},\Delta) \right\}}\\
&& \approx |\vec{E}_{in}(0,0)|^{2} |\vec{E}_{in}(\Delta,\Delta)|^{2} \Bigg[ 1 - \Omega^{2} \Gs \bigg\{ 2 \zeta_{1}(0,\Delta)  - 2 \zeta_{1}(\Delta_{\tau},\Delta) - 2 \zeta_{2}(0,\Delta)  + 2 \zeta_{2}(\Delta_{\tau},\Delta) \bigg\} \Bigg].\\
&& \nonumber \overline{M^{2,2}_{\{\textsc{x,c}\};\{\textsc{x,d}\}} \left( \bm{R};\bm{R} \right)} = \overline{M^{2,2}_{\{\textsc{x,d}\};\{\textsc{x,c}\}} \left( \bm{R};\bm{R} \right)} = \overline{M^{2,2}_{\{\textsc{c,x}\};\{\textsc{d,x}\}} \left( \bm{R};\bm{R} \right)} = \overline{M^{2,2}_{\{\textsc{d,x}\};\{\textsc{c,x}\}} \left( \bm{R};\bm{R} \right)}\\ 
&&= |\vec{E}_{in}(0,0)|^{2} |\vec{E}_{in}(\Delta,\Delta)|^{2} \: e^{- \Omega^{2} \Gs \left\{  \zeta_{1}(0,\Delta)  -  \zeta_{2}(0,\Delta) \right\}} \approx |\vec{E}_{in}(0,0)|^{2} |\vec{E}_{in}(\Delta,\Delta)|^{2} \Bigg[ 1 - \Omega^{2} \Gs \bigg\{  \zeta_{1}(0,\Delta)  -  \zeta_{2}(0,\Delta) \bigg\} \Bigg]
\label{eqn:M22DCMom}
\end{eqnarray}

The two-point correlation of output power is given by~\cite{Shar2024}
\be
\overline{P_{\mathrm{out}}^{i}(\tau) P_{\mathrm{out}}^{j}(\tau+\Delta_{\tau})}
= \left(\epsilon_{0} c \right)^{2} \iint\limits_A  \!\! \mathrm{d}^{2}a_{1} \: \mathrm{d}^{2}a_{2} \: \overline{M^{2,2}_{\mathrm{out}} \left(\bm{R};\bm{R}\right)}.
\label{eqn:P2ptCorr}
\ee
Here Eq. \eref{eqn:P2ptCorr} can be used for any beam with cross-section area $A$. However, in our case, we use the assumption that the width of the light beams is effectively zero. In other words, the light beams have been approximated to light rays. Specifically, in Eq. \eref{eqn:P2ptCorr}, we have used this assumption to simplify the surface integrals to $A^{2}$. Further, we also use input power $P_{0}=2\, \epsilon_{0}\, c\, A\, |\vec{E}_{in}|^{2}$ to allow further simplification that leads to Eq. \eref{eqn:CovSimp} in terms of $P_{0}$.
Here the covariance of output power is
\begin{eqnarray}
\nonumber &&\mathrm{Cov}_{i,j}(P_{\mathrm{out}})=\overline{P^{i}_{\mathrm{out}}(\tau) P^{j}_{\mathrm{out}}(\tau+\Delta_{\tau})}\\
&& \hspace*{8 em}  \, -\overline{P^{i}_{\mathrm{out}}(\tau)} \:\:\: \overline{P^{j}_{\mathrm{out}}(\tau+\Delta_{\tau})}.
\label{eqn:Covij}
\end{eqnarray}
Note that $\mathrm{Cov}_{i,j}(P_{\mathrm{out}}) \neq \mathrm{Cov}_{j,i}(P_{\mathrm{out}})$, in general.

The covariance using Eqs. \eref{eqn:M22DCMom} and \eref{eqn:m22DC} for input power $P_{0}$,
\begin{align}
\nonumber \text{Cov}_{i,j}(P_{\text{out}})= \frac{P_{0}^{2}}{4} &\Bigg[\frac{1}{2} \Bigg( 1 - \Omega^{2} \Gs \bigg\{ 2 \zeta_{1}(0,\Delta)  + 2 \zeta_{1}(\Delta_{\tau},\Delta) - 2 \zeta_{2}(0,\Delta)  - 2 \zeta_{2}(\Delta_{\tau},\Delta) \bigg\}\Bigg) \cos 4 \varphi_{\text{off}}\\
\nonumber & + \frac{1}{2} \Bigg( 1 - \Omega^{2} \Gs \bigg\{ 2 \zeta_{1}(0,\Delta)  - 2 \zeta_{1}(\Delta_{\tau},\Delta) - 2 \zeta_{2}(0,\Delta)  + 2 \zeta_{2}(\Delta_{\tau},\Delta) \bigg\} \Bigg) \\
&- \Bigg( 1 - \Omega^{2} \Gs \bigg\{ 2 \zeta_{1}(0,\Delta)  - 2 \zeta_{2}(0,\Delta) \bigg\} \Bigg) \cos^{2} 2 \varphi_{\text{off}} \Bigg].
\label{eqn:CovSimp}
\end{align}

We know that 
\begin{align}
\text{Cov}_{i,j}(\Delta x)= \left( \frac{\lambda}{4 \pi \varphi_{\text{off}} P_{0}} \right)^{2}\text{Cov}_{i,j}(P_{\text{out}}).
\end{align}
Using trigonometric identities, 
\begin{align}
\nonumber \text{Cov}_{i,j}(\Delta x)= \left( \frac{\lambda}{4 \pi \varphi_{\text{off}} P_{0}} \right)^{2} \frac{P_{0}^{2}}{4} &\Bigg[ \frac{1}{2}\Bigg( 1 - \Omega^{2} \Gs \bigg\{ 2 \zeta_{1}(0,\Delta)  + 2 \zeta_{1}(\Delta_{\tau},\Delta) - 2 \zeta_{2}(0,\Delta)  - 2 \zeta_{2}(\Delta_{\tau},\Delta) \bigg\}\Bigg) (1 - 2 \sin^{2} 2 \varphi_{\text{off}})\\
\nonumber & + \frac{1}{2} \Bigg( 1 - \Omega^{2} \Gs \bigg\{ 2 \zeta_{1}(0,\Delta)  - 2 \zeta_{1}(\Delta_{\tau},\Delta) - 2 \zeta_{2}(0,\Delta)  + 2 \zeta_{2}(\Delta_{\tau},\Delta) \bigg\} \Bigg) \\
&- \Bigg( 1 - \Omega^{2} \Gs \bigg\{ 2 \zeta_{1}(0,\Delta)  - 2 \zeta_{2}(0,\Delta) \bigg\} \Bigg) (1 - \sin^{2} 2 \varphi_{\text{off}}) \Bigg]\\
&= \left( \frac{\lambda \Omega}{4 \pi  } \right)^{2}  \frac{\sin^{2} 2 \varphi_{\text{off}}}{4 \varphi^{2}_{\text{off}}} \Gs \Bigg[ 2 \zeta_{1}(\Delta_{\tau},\Delta)  - 2 \zeta_{2}(\Delta_{\tau},\Delta) \Bigg].
\end{align}
Using $\sin^{2} 2 \varphi_{\text{off}} \approx 4 \varphi^{2}_{\text{off}}$ as $\varphi_{\text{off}} \ll 1$,
\begin{align}
\text{Cov}_{i,j}(\Delta x)= \frac{c^{2} \Gs }{2} &\Bigg[\zeta_{1}(\Delta_{\tau},\Delta) - \zeta_{2}(\Delta_{\tau},\Delta) \bigg].
\label{eqn:CovPathHolo}
\end{align}
We need the optical path difference to be stationary to apply the Wiener-Khinchin theorem for obtaining the spectral densities. For each correlation function, we check if the obtained autocorrelation is non-negative definite to check for weak stationarity. We then obtain the PSD using a cosine transform of $\text{Cov}_{i,i}(\Delta x)$ (setting $\Delta=0$, $\zeta_{1}\to\sigma$ and $\zeta_{2}\to\xi$) while we obtain the CSD using a Fourier transform of the $\text{Cov}_{\textsc{i,ii}}(\Delta x)$ (setting $\Delta=\Delta_{s}$). 

We simplify the PSD expression as follows. We consider 
\begin{align*}
\sigma(\Delta_{\tau})=& \int_{0}^{\Li/c} \rmd t_{1} \int_{0}^{\Li/c} \rmd t_{2} \bigg[\rho((t_{1}-t_{2}+\Delta_{\tau}),c(t_{1}-t_{2}),0,0)\\
&+ \rho\left(\left(\frac{2 \Li}{c} - t_{1}-t_{2}+\Delta_{\tau}\right),c(t_{1}-t_{2}),0,0\right)+ \rho\left(\left(t_{1}+t_{2}-\frac{2 \Li}{c} +\Delta_{\tau}\right),c(t_{1}-t_{2}),0,0\right) \\
&+ \rho((t_{2}-t_{1}+\Delta_{\tau}),c(t_{1}-t_{2}),0,0) \bigg].\\
\xi(\Delta_{\tau})=& \int_{0}^{\Li/c} \rmd t_{1} \int_{0}^{\Li/c} \rmd t_{2} \bigg[\rho((t_{1}-t_{2}+\Delta_{\tau}),c t_{1},0,-c t_{2})\\
&+ \rho\left(\left(\frac{2 \Li}{c} - t_{1}-t_{2}+\Delta_{\tau}\right),c t_{1},0,-c t_{2}\right)+ \rho\left(\left(t_{1}+t_{2}-\frac{2 \Li}{c} +\Delta_{\tau}\right),c t_{1},0,-c t_{2}\right) \\
&+ \rho((t_{2}-t_{1}+\Delta_{\tau}),c t_{1},0,-c t_{2}) \bigg].
\end{align*}
Applying the cosine transform first over each of the four terms of the two correlation functions and using trignometric identities, we obtain a simplified expression of the PSD. We rewrite this simplified PSD $S(f)$ as $S(\nu)$ in terms of $\nu=\pi f \Li/c=\pi f/(2 \fsr)$ using $\vec{\Delta}_{w}=\left(0,0,\Li \, (u_{1}-u_{2})\right)$ and $\vec{\Delta}_{c}=\left(\Li \, u_{1},0,- \Li \, u_{2}\right)$, as
\begin{align}
 S(\nu) &= \frac{2 \Gs \Li^{2}}{\pi} \int\limits_{0}^{1} \rmd u_{1}  \int\limits_{0}^{1} \rmd u_{2} \, \cos\left(2 \nu \left( 1 - u_{1} \right) \right) \cos\left(2 \nu \left( 1 - u_{2} \right) \right) \int\limits_{0}^{\infty} \rmd T \cos \left(2 \nu \frac{c T}{\Li}\right) \left(\rho(c T,\vec{\Delta}_{w}) - \rho(c T,\vec{\Delta}_{c}) \right).
\label{eqn:PSDsimp}
\end{align}

\section{III. Holometer-type setup: Response functions}

In this section, we obtain the interferometer response function for a Holometer-type setup.

We describe an effective phase difference between the two arms of an MLI, given by, 
\begin{equation}
\Delta \Phi_{\textsc{i}}\left(\tau_{0}+\Delta_{\tau},\Delta\right) =  \varphi_{\text{off}} + \Omega \int_{\Delta_{\tau}}^{\Delta_{\tau}+\tau_{0}} \rmd t' \Bigg[ w(t',s(t'-\Delta_{\tau})+\Delta,\Delta,\Delta) - w(t',\Delta,\Delta,s(t'-\Delta_{\tau})+\Delta) \Bigg].
\label{eqn:PhDiff}
\end{equation}
We can verify that this effective phase difference gives Eq. \eref{eqn:CovPathHolo} multiplied by a factor of $(4 \pi/\lambda)^{2}$, on computing $\text{Cov}_{i,j}(\Delta \Phi_{\textsc{i}})= \overline{\Delta\Phi_{\textsc{i}}\left(\tau_{0},0\right) \Delta\Phi_{\textsc{i}}\left(\tau_{0}+\Delta_{\tau},\Delta\right)} -\, \overline{\Delta \Phi_{\textsc{i}}\left(\tau_{0},0\right)} \hspace{0.6 em} \overline{\Delta \Phi_{\textsc{i}}\left(\tau_{0}+\Delta_{\tau},\Delta\right)}$. Further, by using this effective phase difference, we implicitly assume Attributes \ref{mod:metric}-\ref{mod:smallFluct} and Assumptions \ref{Ass:eikonal}-\ref{Ass:iso} listed in Secs. I and II, used in obtaining Eq. \eref{eqn:CovPathHolo}.

Defining the transform,
\begin{align}
w(\bm{r})=w(t,\vec{r}(t))=\int \rmd^{3} \kvec \int_{-\infty}^{\infty} \rmd \omega_{1} \, \widetilde{w}(\omega_{1},\kvec) \, e^{i \left(\omega_{1} t + \kvec\cdot\vec{r}\right)},
\label{eqn:trans}
\end{align}
we rewrite Eq. \eref{eqn:PhDiff} as
\begin{align}
\nonumber \Delta \Phi_{\textsc{i}}\left(\tau_{0}+\Delta_{\tau},\Delta\right) =  \varphi_{\text{off}} + \Omega \int \rmd^{3} \kvec \int_{-\infty}^{\infty} \rmd \omega_{1} &\, \widetilde{w}(\omega_{1},\kvec) \, e^{i \kvec\cdot\vec{\Delta}_{r}} \\
& \int_{\Delta_{\tau}}^{\Delta_{\tau}+\tau_{0}} \rmd t' \Bigg[ e^{i \left(\omega_{1} t' + \kvec\cdot\hat{e}_{x}  s(t'-\Delta_{\tau})\right)} - e^{i \left(\omega_{1} t' + \kvec\cdot\hat{e}_{z}  s(t'-\Delta_{\tau})\right)} \Bigg].
\end{align}
Here $\vec{\Delta}_{r}=(\Delta,\Delta,\Delta)$.
Using the definition of $s(t)$,
\begin{align*}
& \int_{\Delta_{\tau}}^{\Delta_{\tau}+\tau_{0}} \rmd t' e^{i \left(\omega_{1} t' + \kvec\cdot\hat{e}_{x}  s(t'-\Delta_{\tau})\right)} =  e^{i \omega_{1} \Delta_{\tau}} \left[\left(\frac{e^{i (\omega_{1}+ c \kvec\cdot\hat{e}_{x}) \frac{\tau_{0}}{2}}- 1}{i (\omega_{1}+ c \kvec\cdot\hat{e}_{x})} \right) + \left(\frac{e^{\frac{i \omega_{1} \tau_{0}}{2}}-e^{i (\omega_{1}+ c \kvec\cdot\hat{e}_{x}) \frac{\tau_{0}}{2}}}{i (\omega_{1}- c \kvec\cdot\hat{e}_{x})} \right) \right].
\end{align*}
This implies that
\begin{align}
\nonumber \Delta \Phi_{\textsc{i}}\left(\tau_{0}+\Delta_{\tau},\Delta\right) &= \varphi_{\text{off}} + \frac{\Omega \Li}{c} \int \rmd^{3} \kvec \int_{-\infty}^{\infty} \rmd \omega_{1} \, \widetilde{w}(\omega_{1},\kvec) \,  e^{i \kvec\cdot\vec{\Delta}_{r}} e^{i \omega_{1} \Delta_{\tau}} \\
\nonumber &\left[e^{i \frac{\Li}{2 c} (\omega_{1}+ c \kvec\cdot\hat{e}_{x})} \left\{ \text{Sinc}\left(\frac{\Li}{2 c} (\omega_{1}+ c \kvec\cdot\hat{e}_{x})\right) + e^{i \omega_{1} \Li/c} \text{Sinc}\left(\frac{\Li}{2 c} (\omega_{1}-c \kvec\cdot\hat{e}_{x})\right)  \right\}\right.\\
& \left. - e^{i \frac{\Li}{2 c} (\omega_{1}+ c \kvec\cdot\hat{e}_{z})} \left\{ \text{Sinc}\left(\frac{\Li}{2 c} (\omega_{1}+ c \kvec\cdot\hat{e}_{z})\right) + e^{i \omega_{1} \Li/c} \text{Sinc}\left(\frac{\Li}{2 c} (\omega_{1}-c \kvec\cdot\hat{e}_{z})\right)  \right\} \right]
\end{align}

To obtain $\text{Cov}_{i,j}(\Delta \Phi_{\textsc{i}})$, we find using Eq. \eref{eqn:trans} that 
\begin{align}
&\overline{\widetilde{w}}=0 \quad \text{because } \overline{w}=0.\\
&\overline{w(t_{1},\vec{r_{1}}) w(t_{2},\vec{r_{2}})} = \int \rmd^{3} \kvec \int_{-\infty}^{\infty} \rmd \omega_{1} \,  \int \rmd^{3} \vec{k_{2}} \int_{-\infty}^{\infty} \rmd \omega_{2} \, \overline{\widetilde{w}(\omega_{1},\kvec) \, \widetilde{w}(\omega_{2},\vec{k_{2}})} \, e^{i \left(\omega_{1} t_{1} + \kvec\cdot\vec{r_{1}}\right)} e^{i \left(\omega_{2} t_{2} + \vec{k_{2}}\cdot\vec{r_{2}}\right)}, \label{eqn:2pt}
\end{align}
Here, in Eq. \eref{eqn:2pt}, due to stationarity ($\overline{w(t_{1},\vec{r_{1}}) w(t_{2},\vec{r_{2}})}$ is only a function of $\vec{r_{1}}-\vec{r_{2}}$ and $t_{1}-t_{2}$) we require,
\begin{align}
\overline{\widetilde{w}(\omega_{1},\kvec) \, \widetilde{w}(\omega_{2},\vec{k_{2}})}= \Gs \widetilde{\rho}(\omega_{1},\kvec) \delta(\omega_{1}+\omega_{2}) \delta^{(3)}(\kvec+\vec{k_{2}}).
\label{eqn:CorrTrans}
\end{align}

Therefore, the covariance of the phase difference becomes
\begin{align}
\nonumber \text{Cov}_{i,j}(\Delta \Phi_{\textsc{i}}) &= \Gs \left(\frac{\Omega \Li}{c}\right)^{2} \int \rmd^{3} \kvec \int_{-\infty}^{\infty} \rmd \omega_{1} \, \widetilde{\rho}(\omega_{1},\kvec) \, e^{i \kvec\cdot\vec{\Delta}_{r}} \\
\nonumber & \left\vert e^{i \frac{\Li}{2 c} (\omega_{1}+ c \kvec\cdot\hat{e}_{x})} \left\{ \text{Sinc}\left(\frac{\Li}{2 c} (\omega_{1}+ c \kvec\cdot\hat{e}_{x})\right) + e^{i \omega_{1} \Li/c} \text{Sinc}\left(\frac{\Li}{2 c} (\omega_{1}-c \kvec\cdot\hat{e}_{x})\right)  \right\}\right.\\
& \left. - e^{i \frac{\Li}{2 c} (\omega_{1}+ c \kvec\cdot\hat{e}_{z})} \left\{ \text{Sinc}\left(\frac{\Li}{2 c} (\omega_{1}+ c \kvec\cdot\hat{e}_{z})\right) + e^{i \omega_{1} \Li/c} \text{Sinc}\left(\frac{\Li}{2 c} (\omega_{1}-c \kvec\cdot\hat{e}_{z})\right)  \right\}  \right\vert^{2}  e^{i \omega_{1} \Delta_{\tau}}.
\end{align}
The covariance of the optical path difference is
\begin{align}
\nonumber \text{Cov}_{i,j}(\Delta x) &= \left(\frac{\lambda}{4 \pi} \right)^{2} \text{Cov}_{i,j}(\Delta \Phi_{\textsc{i}})\\
\nonumber & = \Gs \left(\frac{\Li}{2}\right)^{2} \int \rmd^{3} \kvec \int_{-\infty}^{\infty} \rmd \omega_{1} \, \widetilde{\rho}(\omega_{1},\kvec) \, e^{i \kvec\cdot\vec{\Delta}_{r}} \\
\nonumber & \left\vert e^{i \frac{\Li}{2 c} (\omega_{1}+ c \kvec\cdot\hat{e}_{x})} \left\{ \text{Sinc}\left(\frac{\Li}{2 c} (\omega_{1}+ c \kvec\cdot\hat{e}_{x})\right) + e^{i \omega_{1} \Li/c} \text{Sinc}\left(\frac{\Li}{2 c} (\omega_{1}-c \kvec\cdot\hat{e}_{x})\right)  \right\}\right.\\
& \left. - e^{i \frac{\Li}{2 c} (\omega_{1}+ c \kvec\cdot\hat{e}_{z})} \left\{ \text{Sinc}\left(\frac{\Li}{2 c} (\omega_{1}+ c \kvec\cdot\hat{e}_{z})\right) + e^{i \omega_{1} \Li/c} \text{Sinc}\left(\frac{\Li}{2 c} (\omega_{1}-c \kvec\cdot\hat{e}_{z})\right)  \right\}  \right\vert^{2}  e^{i \omega_{1} \Delta_{\tau}}.
\end{align}
The corresponding power spectral density (PSD) with $\Delta=0$ in $\vec{\Delta}_{r}=(\Delta,\Delta,\Delta)$ is
\begin{align}
S(f)&= \frac{1}{2 \pi} \int_{-\infty}^{\infty} \rmd \Delta_{\tau} e^{-2 \pi i f \Delta_{\tau}} \text{Cov}_{i,i}(\Delta x)\\
\nonumber & = \Gs \left(\frac{\Li}{2}\right)^{2} \int \rmd^{3} \kvec \: \widetilde{\rho}(2 \pi f,\kvec) \, \left\vert e^{i \frac{\Li}{2 c} (2 \pi f+ c \kvec\cdot\hat{e}_{x})} \left\{ \text{Sinc}\left(\frac{\Li}{2 c} (2 \pi f+ c \kvec\cdot\hat{e}_{x})\right) + e^{i 2 \pi f \Li/c} \text{Sinc}\left(\frac{\Li}{2 c} (2 \pi f-c \kvec\cdot\hat{e}_{x})\right)  \right\}\right.\\
& \left. - e^{i \frac{\Li}{2 c} (2 \pi f+ c \kvec\cdot\hat{e}_{z})} \left\{ \text{Sinc}\left(\frac{\Li}{2 c} (2 \pi f+ c \kvec\cdot\hat{e}_{z})\right) + e^{i 2 \pi f \Li/c} \text{Sinc}\left(\frac{\Li}{2 c} (2 \pi f-c \kvec\cdot\hat{e}_{z})\right)  \right\}  \right\vert^{2}.
\end{align}
The correponding cross spectral density (CSD) with a non-zero $\Delta$ is
\begin{align}
CS(f)&= \frac{1}{2 \pi} \int_{-\infty}^{\infty} \rmd \Delta_{\tau} e^{-2 \pi i f \Delta_{\tau}} \text{Cov}_{i.j}(\Delta x)\\
\nonumber & = \Gs \left(\frac{\Li}{2}\right)^{2} \int \rmd^{3} \kvec \: \widetilde{\rho}(2 \pi f,\kvec) \, e^{i \kvec\cdot\vec{\Delta}_{r}} \\
\nonumber & \left\vert e^{i \frac{\Li}{2 c} (2 \pi f+ c \kvec\cdot\hat{e}_{x})} \left\{ \text{Sinc}\left(\frac{\Li}{2 c} (2 \pi f+ c \kvec\cdot\hat{e}_{x})\right) + e^{i 2 \pi f \Li/c} \text{Sinc}\left(\frac{\Li}{2 c} (2 \pi f-c \kvec\cdot\hat{e}_{x})\right)  \right\}\right.\\
& \left. - e^{i \frac{\Li}{2 c} (2 \pi f+ c \kvec\cdot\hat{e}_{z})} \left\{ \text{Sinc}\left(\frac{\Li}{2 c} (2 \pi f+ c \kvec\cdot\hat{e}_{z})\right) + e^{i 2 \pi f \Li/c} \text{Sinc}\left(\frac{\Li}{2 c} (2 \pi f-c \kvec\cdot\hat{e}_{z})\right)  \right\}  \right\vert^{2}.
\end{align}
We note here that by setting $\Delta=0$ in $\vec{\Delta}_{r}$, we recover $S(f)$.

Considering that the interferometer response function $\widetilde{\chi}_{\Delta}(f,\kvec)$ corresponding to the CSD, is defined using
\begin{align}
CS(f) &= \int \rmd^{3} \kvec \: \Gs \, \widetilde{\rho}(2 \pi f,\kvec) \, \widetilde{\chi}_{\Delta}(f,\kvec), 
\end{align}
we find
\begin{align}
\nonumber \widetilde{\chi}_{\Delta}(f,\kvec)&= \left(\frac{\Li}{2}\right)^{2} \, e^{i \kvec\cdot\vec{\Delta}_{r}} \, \left\vert e^{i \frac{\Li}{2 c} (2 \pi f+ c \kvec\cdot\hat{e}_{x})} \left\{ \text{Sinc}\left(\frac{\Li}{2 c} (2 \pi f+ c \kvec\cdot\hat{e}_{x})\right) + e^{i 2 \pi f \Li/c} \text{Sinc}\left(\frac{\Li}{2 c} (2 \pi f-c \kvec\cdot\hat{e}_{x})\right)  \right\}\right.\\
& \left. - e^{i \frac{\Li}{2 c} (2 \pi f+ c \kvec\cdot\hat{e}_{z})} \left\{ \text{Sinc}\left(\frac{\Li}{2 c} (2 \pi f+ c \kvec\cdot\hat{e}_{z})\right) + e^{i 2 \pi f \Li/c} \text{Sinc}\left(\frac{\Li}{2 c} (2 \pi f-c \kvec\cdot\hat{e}_{z})\right)  \right\}  \right\vert^{2}.
\end{align}
This is rewritten as
\begin{align}
&\widetilde{\chi}_{\Delta}(f,\kvec)= \left(\frac{\Li}{2}\right)^{2} \, e^{i \kvec\cdot\vec{\Delta}_{r}} \, \left\vert C_{x}(f,\kvec) - C_{z}(f,\kvec) \right\vert^{2},
\end{align}
with 
\begin{align}
C_{j}(f,\kvec)&=e^{i f T_{+}^{(j)}} \left\{ \text{Sinc}\left(f T_{+}^{(j)}\right) + e^{\frac{2 \pi i f \Li}{c}} \text{Sinc}\left(f T_{-}^{(j)} \right)  \right\}, \label{eq:HoloRFpart}\\
T_{\pm}^{(j)}(f,\kvec)&=\frac{\pi \Li}{c} \left(1 \pm \frac{c}{2 \pi f} \, \kvec\cdot\hat{e}_{j}\right), \quad (j=x,z).
\end{align}
We note that the interferometer response function $\widetilde{\chi}_{0}(f,\kvec)$ corresponding to $\Delta=0$ for obtaining PSD $S(f)$, is denoted simply by $\widetilde{\chi}_{\textsc{i}}(f,\kvec)$ for ease of notation. Here the PSD is then given by
\begin{align}
S(f) &= \int \rmd^{3} \kvec \: \Gs \, \widetilde{\rho}(2 \pi f,\kvec) \, \widetilde{\chi}_{\textsc{i}}(f,\kvec), 
\end{align}
with 
\begin{align}
&\widetilde{\chi}_{\textsc{i}}(f,\kvec)= \left(\frac{\Li}{2}\right)^{2} \, \left\vert C_{x}(f,\kvec) - C_{z}(f,\kvec) \right\vert^{2}.
\label{eqn:InterRF}
\end{align}

\section{IV. LIGO: Spectral densities and response functions}

In this section, we present the signal PSD of the opyical path difference between the two arms of LIGO setup. We also present the interferometer response function in this setup.

LIGO is an MLI with cavities in each arm~\cite{LIGOdata} as shown in Fig. \ref{fig:aLIGOsetup}. These arm cavities are formed by introducing a mirror in each arm.
The electric field at the detector B is 
\begin{eqnarray}
 &E_{\mathrm{out}}(\tau ,0,0,0) = \frac{1}{\sqrt{2}} \sum\limits_{q=1}^{\infty} T_{\textsc{m}}\, \sqrt{ R_{\textsc{m}}^{q-1}} \, \bigg[ E^{(\textsc{c})}_{y}(\tau ,0,0,0) - E^{(\textsc{d})}_{y}(\tau ,0,0,0) e^{- 2 i \varphi_{\text{off}}} \bigg].
 \label{eqn:EoutLIGO}
\end{eqnarray}
Here $T_{\textsc{m}}=1-R_{\textsc{m}}$ 
is the transmission coefficient of the mirrors introduced to render arm cavities. Equation \eref{eqn:Eout} describes the corresponding output electric field in the Holometer-type setup. We use $T_{\textsc{m}}=0.014,$ whereby $R_{\textsc{m}}^{280} \approx 0.019 < 0.02$, 
i.e., less than 2$\%$ of the input light remains after $280$ round-trips of the light beam (i.e., finesse of LIGO setup) in each arm.

Using Eq. \eref{eqn:EoutLIGO} in place of Eq. \eref{eqn:Eout} and implementing the procedure described in Sec. II (assuming Attributes \ref{mod:metric}-\ref{mod:smallFluct} and Assumptions \ref{Ass:eikonal}-\ref{Ass:iso}), we find the PSD to be 
\begin{align}
S(f)=& \frac{c^{2} \Gs T_{\textsc{m}}^{4}}{2} \left( \frac{1}{1- \sqrt{R_{\textsc{m}}}} \right)^{2} \sum\limits_{q_{1},q_{2}=1}^{\infty} (\sqrt{R_{\textsc{m}}})^{q_{1}+q_{2}-2} \, \int_{0}^{\infty} \rmd \Delta_{\tau}  \Bigg[\sigma^{(q_{1},q_{2})}(\Delta_{\tau}) - \xi^{(q_{1},q_{2})}(\Delta_{\tau}) \bigg] \cos 2 \pi f \Delta_{\tau},
\end{align}
where
\begin{eqnarray}
&&\sigma^{(p,q)}(\Delta_{\tau})= \int\limits_{0}^{p\tau_{0}} \rmd t_{1} \int\limits_{0}^{q\tau_{0}} \rmd t_{2} \, \rho(t_{1}+\Delta_{\tau}-t_{2},0,0,s(t_{1})-s(t_{2})) = \int\limits_{0}^{p\tau_{0}} \rmd t_{1} \int\limits_{0}^{q\tau_{0}} \rmd t_{2} \, \rho(t_{1}+\Delta_{\tau}-t_{2},s(t_{1})-s(t_{2}),0,0),
\label{eqn:SigLIGODefn}\\
&&\xi^{(p,q)}(\Delta_{\tau}) = \int\limits_{0}^{p\tau_{0}} \rmd t_{1} \, \int\limits_{0}^{q\tau_{0}} \rmd t_{2} \, \rho(t_{1}+\Delta_{\tau}-t_{2},s(t_{1}),0,-s(t_{2})).\label{eqn:XiLIGODefn}
\end{eqnarray}
Here $s(t)=c t - 2 q \Li$ if $q \tau_{0} \leqslant t \leqslant \left(\frac{2 q + 1}{2}\right) \tau_{0}$ and $s(t)=2 (q+1) \Li - c t$ if $\left(\frac{2 q + 1}{2}\right) \tau_{0}<t\leqslant (q+1) \tau_{0}$.\\

\begin{figure*}
\begin{tikzpicture}
\draw[red,thick] (0.,0.05) -- (1.95,0.05);
\draw[red,thick] (2.,0.) -- (4.,0.);
\draw[red,thick] (2.,-2.) -- (2.,0.);
\draw[red,thick] (1.95,0.05) -- (1.95,2.);
\draw[red,thick,->] (0.,0.05) -- (1.,0.05);
\draw[red,thick,->] (2.,0.) -- (3.,0.);
\draw[red,thick,->] (1.95,0.05) -- (1.5,0.05);
\draw[red,thick,->] (4.,0.) -- (3.5,0.);
\draw[red,thick,->] (2.,0.) -- (2.,-1.5);
\draw[red,thick,->] (1.95,0.05) -- (1.95,1.);
\draw[red,thick,->] (1.95,2.) -- (1.95,1.5);
\draw[black,thick] (4.,-0.2) -- (4.,0.2);
\draw[black,thick] (1.75,2.) -- (2.15,2.);
\draw[black,thick] (1.85,-0.2) node[left] (scrip) {BS};
\draw[black!30,fill=black!30] (1.8,-0.2) -- (2.2,0.2)--(2.15,0.25)--(1.75,-0.15)--cycle;
\draw[black,thick] (1.8,-0.2) -- (2.2,0.2);
\draw[black,thick,<->] (2.1,-0.5) -- (2.5,-0.5);
\draw[black,thick] (2.1,-0.7) node[right] (scrip) {\scriptsize
 $d_{\textsc{d}}$};
\draw[black,thick,<-] (2.7,-0.5) -- (3.1,-0.5) node[right] (scrip) {\scriptsize
$\Li$};
\draw[black,thick,->] (3.55,-0.5) -- (3.9,-0.5);
\draw[black,thick,<->] (1.5,0.1) -- (1.5,0.55);
\draw[black,thick] (1.5,0.35) node[left] (scrip) {\scriptsize 
$d_{\textsc{c}}$};
\draw[black,thick,<-] (1.5,0.7) -- (1.5,1.1) node[left] (scrip) {\scriptsize
 $\Li$};
\draw[black,thick,->] (1.5,0.9) -- (1.5,1.9);
\draw[black,fill=black!30] (-0.1,-0.1) rectangle (0.5,0.2);
\draw[lasernode] (-0.05,-0.05) rectangle (0.45,0.15);
\draw[black,thick,fill=red] (2.2,-1.8) -- (1.8,-1.8) arc(180:360:0.2) --cycle;
\draw[black,thick,snake it] (2,-2) -- (2.1,-2.5);
\draw[black,thick] (0.,0.05) node[left] (scrip) {\scriptsize
A};
\draw[black,thick] (1.9,-2.) node[left] (scrip) {\scriptsize
B};
\draw[black,thick] (4.,0.) node[right] (scrip) {\scriptsize
D};
\draw[black,thick] (2.,2.2) node[right] (scrip) {\scriptsize
C};
\draw[black,thick,->] (0.2,-1.8) -- (0.5,-1.8) node[right] (scrip) {\scriptsize
 z};
\draw[black,thick,->] (0.2,-1.8) -- (0.2,-1.5) node[left] (scrip) {\scriptsize
 x};
\draw[black,thick,->] (0.2,-1.8) -- (0.1,-2) node[left] (scrip) {\scriptsize
 y};
\draw[black, thick] (1.75,0.6) -- (2.15,0.6);
\draw[black, thick] (2.6,-0.2) -- (2.6,0.2);
\draw[black,thick] (-0.5,2.) node[left] (scrip) {\scriptsize
(a)};
\end{tikzpicture}
\begin{tikzpicture}
\draw[black,thick] (1.,0.8) node[right] (scrip) {\scriptsize (b)};
\draw[red,thick] (2.,0.) -- (6.,0.);
\draw[red,thick] (2.,0.) -- (2.,-0.4);
\draw[red,thick,->] (6.,0.) -- (5.,0.);
\draw[red,thick,->] (2.,0.) -- (4.5,0.);
\draw[black,thick] (6.,-0.4) -- (6.,0.4);
\draw[black,thick] (1.65,-0.4) node[left] (scrip) {\scriptsize BS};
\draw[black!30,fill=black!30] (1.6,-0.4) -- (2.4,0.4)--(2.3,0.5)--(1.5,-0.3)--cycle;
\draw[black,thick] (1.6,-0.4) -- (2.4,0.4);
\draw[black, thick] (3.5,-0.4) node[below] (scrip) {\scriptsize M} -- (3.5,0.4);
\draw[green!40!black,thick,<-] (2.9,0.6)node[above] (scrip) {\scriptsize $T_{\textsc{m}}$} -- (3.4,0.6);
\draw[green!40!black,thick,->] (3.5,0.6) -- (4.,0.6) node[above] (scrip) {\scriptsize $R_{\textsc{m}}$};
\draw[black,thick,<->] (2.1,-1) -- (3.4,-1);
\draw[black,thick] (2.5,-1.2) node[right] (scrip) {\scriptsize $d_{\textsc{d}}$};
\draw[black,thick,<-] (3.6,-1) -- (4.5,-1) node[right] (scrip) {\scriptsize $\Li$};
\draw[black,thick,->] (5,-1) -- (5.9,-1);
\draw[black,thick] (6.,0.) node[right] (scrip) {\scriptsize D};
\end{tikzpicture}
\caption{(a) MLI with arm cavities and $d_{\textsc{d}}-d_{\textsc{c}}=\varphi_{\text{off}}/k$, and (b) arm D of the interferometer.}
\label{fig:aLIGOsetup}
\end{figure*}
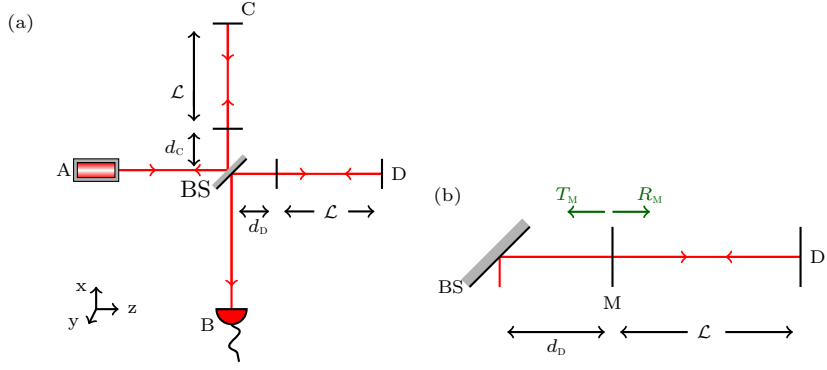

To obtain the response function in the case of aLIGO, we use the effective phase difference,
\begin{align}
\nonumber \Delta \Phi_{\textsc{l}}\left(\tau_{0}+\Delta_{\tau}\right) =  \varphi_{\text{off}} + \Omega T_{\textsc{m}}^{2} \left( \frac{1}{1- \sqrt{R_{\textsc{m}}}} \right) \sum_{q=1}^{\infty} &\left( \sqrt{R_{\textsc{m}}} \right)^{q-1} \\
&\int_{\Delta_{\tau}}^{\Delta_{\tau}+q \tau_{0}} \rmd t' \Bigg[ w(t',s(t'-\Delta_{\tau}),0,0) - w(t',0,0,s(t'-\Delta_{\tau})) \Bigg].
\end{align} 
By using this effective phase difference, we implicitly assume Attributes \ref{mod:metric}-\ref{mod:smallFluct} and Assumptions \ref{Ass:eikonal}-\ref{Ass:iso} listed in Secs. I and II. We reiterate that these assumptions include all correlation functions that model isotropic, Gaussian spacetime fluctuations (SFs). Implementing the procedure described in Sec. III for the above phase difference we obtain
\begin{align}
S(f) &= \int \rmd^{3} \kvec \: \Gs \, \widetilde{\rho}(2 \pi f,\kvec) \, \widetilde{\chi}_{\textsc{l}}(f,\kvec), 
\end{align}
we find
\begin{align}
&\widetilde{\chi}_{\textsc{l}}(f,\kvec)= \widetilde{\chi}_{\textsc{i}}(f,\kvec) \, \widetilde{\chi}_{\textsc{fp}}(f,\kvec), \label{eq:RFaLIGO}\\
&\widetilde{\chi}_{\textsc{fp}}(f,\kvec)= T_{\textsc{m}}^{4} \left( \frac{1}{1- \sqrt{R_{\textsc{m}}}} \right)^{4} \, \left( \frac{1}{1+ R_{\textsc{m}}- 2 \sqrt{R_{\textsc{m}}} \cos (4 \pi  f \Li/c)} \right). \label{eq:RFcav}
\end{align}
Here we recall $\widetilde{\chi}_{\textsc{i}}(f,\kvec)$ from Eq. \eref{eqn:InterRF} and $C_{i}(f,\kvec)$ from Eq. \eref{eq:HoloRFpart},
\begin{align*}
\widetilde{\chi}_{\textsc{i}}(f,\kvec)&= \left(\frac{\Li}{2}\right)^{2} \, \left\vert C_{x}(f,\kvec) - C_{z}(f,\kvec) \right\vert^{2},\\
C_{j}(f,\kvec)&=e^{i f T_{+}^{(j)}} \left\{ \text{Sinc}\left(f T_{+}^{(j)}\right) + e^{\frac{2 \pi i f \Li}{c}} \text{Sinc}\left(f T_{-}^{(j)} \right)  \right\},
\end{align*}
with
\begin{align*}
T_{\pm}^{(j)}(f,\kvec)&=\frac{\pi \Li}{c} \left(1 \pm \frac{c}{2 \pi f} \, \kvec\cdot\hat{e}_{j}\right), \quad (j=x,z).
\end{align*}
It is evident from Eq.~\eref{eq:RFaLIGO} that the response function of LIGO setup factorises into the response functions of the Fabry-P\'erot arm cavity and a simple MLI without arm cavities. It is also evident from the above expression that the light-crossing frequency $\fsr=c/(2 \Li)$ is the most dominant frequency scale. We point out that this gain computed, while identical in features to the one obtained in \cite[Eq. (A20)]{cahillane22}, has minor differences due to the following two reasons: (1) we consider interference from two arm cavities instead of a single cavity assumed in Appendix A of \cite{cahillane22} and (2) we assume perfect reflectivity of the end mirrors at C and D.

In the limit $\nu \ll 1$, using Taylor series expansion, we find
\begin{align}
\widetilde{\chi}_{\textsc{fp}}(f,\kvec)= T_{\textsc{m}}^{4} \left( \frac{1}{1- \sqrt{R_{\textsc{m}}}} \right)^{6} \left( 1- \frac{ 16 \sqrt{R_{\textsc{m}}} \nu^{2}}{(1-\sqrt{R_{\textsc{m}}})^{2}} +\text{O}(\nu^{4}) \right).
\label{eqn:FPgain}
\end{align}
Here $\nu= \frac{\pi f }{2 \fsr}$. For $\nu \ll 1$, expanding to O$(\nu^{2})$, we can immediately see that $\widetilde{\chi}_{\textsc{fp}}(f,\kvec)$ is inversely proportional to $\nu^{2}$. Further, it is evident that in the limit $\nu\to 0$, the response function of LIGO setup saturates to a product of the gain from the joint effect of the two Fabry-P\'erot arm cavities, $T_{\textsc{m}}^{4} \left( \frac{1}{1- \sqrt{R_{\textsc{m}}}} \right)^{6}$, and response function of a simple MLI. 

%

\end{widetext}

\section{V. Factorised correlation function: PSD and properties}

In this and the following section, we analyse the trends of the PSD in the case of an MLI at low- and high-frequency limits.

For any 3-vector $\DelR$ and time interval $\DelT$, the factorised correlation function $\rho_{\textsc{f}}\left( c \Delta_{t}, \vec{\Delta}_{r} \right)$ is of the form $\rho_{\textsc{f}}\left( c \Delta_{t}, \vec{\Delta}_{r} \right)=\rho_{\text{t}}(c \Delta_{t}) \rho_{\text{s}}(\vec{\Delta}_{r})$ with $\ell_{r}$ setting the correlation scale in space and time. Substituting this into Eq. \eref{eqn:PSDsimp}, it is easily seen that
\begin{align}
S_{\textsc{nc}}(\nu) &= \frac{2}{\pi} \mathcal{S}_{\text{s}}(\nu) \mathcal{S}_{\text{t}}(\nu),\\
\nonumber \text{with } \mathcal{S}_{\text{s}}(\nu) &= \int\limits_{0}^{1} \rmd u_{1}  \int\limits_{0}^{1} \rmd u_{2} \, \cos\left(2 \nu \left( 1 - u_{1} \right) \right)\\
 &\hspace*{-2 em} \cos\left(2 \nu \left( 1 - u_{2} \right)\right) \,\left(\rho_{\text{s}}(\vec{\Delta}_{\parallel}) - \rho_{\text{s}}(\vec{\Delta}_{\perp}) \right),\\
\mathcal{S}_{\text{t}}(\nu) &=\int_{0}^{\infty} \rmd \varphi \: \rho_{\text{t}}(\Li \varphi) \, \cos 2 \nu \varphi.
\end{align}
Here, with $t_{i}=\Li u_{i}/c$ ($i=1,2$), we use $\vec{\Delta}_{\parallel}=(0,0,s(t_{1})-s(t_{2}))$ and $\vec{\Delta}_{\perp}=(s(t_{1}),0,-s(t_{2}))$. We recall $s(t)=c t$ if $0\leqslant t \leqslant \tau_{0}/2$ and $s(t)=2 \Li - c t$ if $\tau_{0}/2<t\leqslant\tau_{0}$. Here we note that the temporal shift $T$ (originating from the covariance definition) in Eq. \eref{eqn:PSDsimp} has been scaled suitably to give a dimensionless $\varphi=c T/\Li$.  

Further, we consider the correlation function $\rho_{\text{s}}$ (respectively, $\rho_{\text{t}}$) to decay with increase in the spatial (respectively, temporal) separation. We find that the above PSD $S_{\textsc{nc}}(\nu)$ is non-zero at $\nu=0$ ($S_{\textsc{nc}}(\nu=0)>0$) and it decays with increase in scaled freqency $\nu$. This can be concluded using the following arguments. Considering that the \textit{vector magnitude} $\Vert \vec{\Delta}_{\parallel} \Vert \leqslant \Vert \vec{\Delta}_{\perp} \Vert$ by the geometry of the interferometer with a negligibly small number of points at which the equality is achieved. Therefore, it is evident that $\rho_{\text{s}}(\vec{\Delta}_{\parallel}) > \rho_{\text{s}}(\vec{\Delta}_{\perp})$ at almost all points. As the other cosine terms in the integral tend to one as $\nu \to 0$, the PSD $S_{\textsc{nc}}(\nu=0)$ is \textit{non-zero, finite} and positive (PSD needs to be positive by definition). Further, as we expect both $\rho_{i}$ ($i$=s,t) to decrease with increase in the corresponding separation, we see that the cosine transforms of such a function will decay with increase in frequency with an appropriate frequency scale.

We can also infer the above from the following mathematical argument. Using the Taylor expansion of the cosine functions in the integrals at $\nu\to0$, we see that 
\begin{align}
\nonumber \mathcal{S}_{\text{s}}(\nu) &= \int\limits_{0}^{1} \rmd u_{1}  \int\limits_{0}^{1} \rmd u_{2} \, \left(\rho_{\text{s}}(\vec{\Delta}_{\parallel}) - \rho_{\text{s}}(\vec{\Delta}_{\perp}) \right)\\
 &\hspace*{-2 em} \left(1- 2 \nu^{2} \left(\left( 1 - u_{1} \right)^{2} + \left( 1 - u_{2} \right)^{2} \right) + \text{O}(\nu^{4}) \right),\\
\mathcal{S}_{\text{t}}(\nu) &=\int_{0}^{\infty} \rmd \varphi \: \rho_{\text{t}}(\Li \varphi) \, (1- 2 \nu^{2} \varphi^{2} + \text{O}(\nu^{4})).
\end{align}
We can see that all combinations to order $\nu^{2}$ shows that the PSD does not increase with increase in frequency at the low-frequency range. 

Considering the fact that the factorised correlation function is assumed to decrease with increase in the separation with \textit{finite} correlation scales, we expect the cosine transforms at high frequencies to decay too. So the resulting PSD has an overall decaying trend, barring any local oscillatory behaviour for any general factorised correlation function. 

In the specific case of the Oppenheim model, we first note that $S_{\textsc{nc}}(\nu)$ is independent of the choice of $\Li$. This is because
\begin{align}
\nonumber S_{\textsc{nc}}(\nu) &= \frac{2}{\pi} \int\limits_{0}^{1} \rmd u_{1}  \int\limits_{0}^{1} \rmd u_{2} \, \cos\left(2 \nu \left( 1 - u_{1} \right) \right)\\
 &\hspace*{-2 em} \cos\left(2 \nu \left( 1 - u_{2} \right)\right) \,\left(\sqrt{u_{1}^{2}+u_{2}^{2}} - | u_{2} - u_{1} | \right),
\end{align} 
is evidently independent of $\Li$. In the limit $\nu\ll 1$, we find numerically that the logarithmic derivative of the PSD with respect to $\nu$, is $-5 \nu/\pi$. For instance, a numerical fit of the PSD in the limit $\nu\ll 1$, yields $S_{\textsc{nc}}(\nu\ll1)\approx 0.275 e^{- \frac{5 \nu^{2}}{2 \pi}}$ for $\Li=3$ m. We also find numerically that the scaled PSD $S_{\textsc{nc}}(\nu)\approx\frac{1}{6 \nu^{2}}$ in the limit $\nu\gg 1$.

\section{VI. Inverse and exponential correlation functions: PSD and properties}

As in the previous section, we examine the PSD at the MLI corresponding to the two classes of correlation functions that cannot be factorised into spatial and temporal parts, at the low- and high-frequency limits.  

\subsection{Inverse functions}
Using Eq. \eref{eqn:PSDsimp}, the scaled PSD corresponding to $\rho_{\textsc{i}{\scriptscriptstyle m}}$ ($m=\textsc{s},\textsc{st}$) is
\begin{align}
\nonumber S_{\textsc{c}}(\nu) &= \frac{2}{\pi} \int\limits_{0}^{1} \rmd u_{1}  \int\limits_{0}^{1} \rmd u_{2} \, \cos\left(2 \nu \left( 1 - u_{1} \right) \right)\\
& \cos\left(2 \nu \left( 1 - u_{2} \right) \right) \left(\mathcal{P}_{\textsc{i}m}(\vec{\Delta}_{\parallel}) - \mathcal{P}_{\textsc{i}m}(\vec{\Delta}_{\perp}) \right).
\label{eqn:PSDinv}
\end{align}
where
\begin{align}
&\mathcal{P}_{\textsc{is}}(\vec{\Delta}_{j})= \, \text{sinc}\left( \nu_{s}^{(j)} \right),
&\mathcal{P}_{\textsc{ist}}(\vec{\Delta}_{j})= \frac{\pi}{2} \, J_{0}\left( \nu_{s}^{(j)} \right),
\label{eqn:TransCorrInv}
\end{align}
with $\nu_{s}^{(j)}= 2 \nu \Vert \vec{\Delta}_{j} \Vert /\Li$ ($j=\parallel,\perp$) and $J_{\alpha}(z)$ being the Bessel function of the first kind. It is useful to note that by virtue of the functional forms of $\mathcal{P}_{\textsc{i}m}$ and Eq. \eref{eqn:PSDinv}, we find $S_{\textsc{c}}(\nu)$ corresponding to $\rho_{\textsc{i}m}$ ($m=\textsc{s,st}$) independent of $\Li$ and $\ell_{r}$.

Using the Taylor expansion of Eq. \eref{eqn:TransCorrInv} about $\nu_{s}^{(j)}=0$ ($j=\parallel,\perp$) in Eq. \eref{eqn:PSDinv} yields the limiting behaviour of $S_{\textsc{c}}(\nu)$ as $\nu\to 0$. 
\begin{align}
&\lim_{\nu\to0} S_{\textsc{c}}(\nu) = c_{\textsc{i}{\scriptscriptstyle m}} \nu^{2}, \quad (m=\textsc{s},\textsc{st}),\\
&c_{\textsc{is}}=2/(3\pi), \quad c_{\textsc{ist}}=1/2.
\end{align}
Here we note that $S_{\textsc{c}}(\nu=0)=0$.

At the high-frequency limit $\nu\gg 1$, it is evident that $\mathcal{P}_{\textsc{is}}$ (Eq. \eref{eqn:TransCorrInv}), which is sinc function, has a $1/\nu$ dependence. This carries forward to the PSD as a $1/\nu$ dependence in this limit. Similarly, for $\rhoist$, the corresponding $\mathcal{P}_{\textsc{ist}}$ (Eq. \eref{eqn:TransCorrInv}) can be approximated in the limit $\nu \gg 1$, as
\begin{align}
\mathcal{P}_{\textsc{ist}}(\vec{\Delta}_{j}) \approx \left. \sin \left(\frac{\pi}{4} + 2 \nu_{s}^{(j)} \right) \middle/ 2 \sqrt{\pi \nu_{s}^{(j)}}.\right.
\end{align}
It is immediately evident that the spacetime-based $\rhoist$ yields a PSD with $1/\sqrt{\nu}$ dependence in this limit.

\subsection{Exponential functions}
The PSD for $\rho_{\textsc{e}{\scriptscriptstyle m}}$ is
\begin{align}
\nonumber S_{\textsc{c}}(\nu) &= \frac{2 \Li}{\pi \ell_{r}} \int\limits_{0}^{1} \rmd u_{1}  \int\limits_{0}^{1} \rmd u_{2} \, \cos\left(2 \nu \left( 1 - u_{1} \right) \right)\\
& \cos\left(2 \nu \left( 1 - u_{2} \right) \right) \left(\mathcal{P}_{\textsc{e}m}(\vec{\Delta}_{\parallel}) - \mathcal{P}_{\textsc{e}m}(\vec{\Delta}_{\perp}) \right).
\label{eqn:PSDexp}
\end{align}
Here
\begin{align}
\label{eqn:TransCorrExp1}
&\mathcal{P}_{\textsc{es}}(\vec{\Delta}_{j})= \, \frac{\Vert \vec{\Delta}_{j} \Vert}{\Li} \, e^{-\frac{\Li}{\ell_{r}} \frac{\Vert \vec{\Delta}_{j} \Vert}{\Li}} \, \text{sinc}\left( \nu_{s}^{(j)} \right),\\
&\mathcal{P}_{\textsc{est}}(\vec{\Delta}_{j})= \int_{0}^{\frac{\Vert \vec{\Delta}_{j} \Vert}{\Li}} \rmd \varphi \: e^{-\frac{\Li}{\ell_{r}} \sqrt{(\Vert \vec{\Delta}_{j} \Vert/\Li)^{2} - \varphi^{2}}} \cos\left(2 \,\nu \, \varphi \right),
\label{eqn:TransCorrExp2}
\end{align}
with $\nu_{s}^{(j)}= 2 \nu \Vert \vec{\Delta}_{j} \Vert /\Li$ ($j=\parallel,\perp$). As before, we have used the dimensionless $\varphi=c T/\Li$ in Eq. \eref{eqn:PSDsimp} to obtain Eq. \eref{eqn:TransCorrExp2} along with substituting in $\rhoest$. We note that $\vec{\Delta}_{j}/\Li$ depends only on the pair $(u_{1},u_{2})$, and is independent of $\Li$. This clearly implies that Eqs. \eref{eqn:TransCorrExp1} and \eref{eqn:TransCorrExp2} depend on  the ratio $\rat=\ell_{r}/\Li$. Therefore, in contrast to the PSDs corresponding to correlation classes considered earlier, $S_{\textsc{c}}(\nu)$ corresponding to $\rho_{\textsc{e}m}$ ($m=\textsc{s,st}$) is dependent on this ratio $\rat$.

\subsubsection{For $\rhoes$} At low frequencies, the Taylor expansion is
\begin{align}
\nonumber & S_{\textsc{c}}(\nu) = \frac{2}{\pi \ell_{r} } \int\limits_{0}^{1} \rmd u_{1}  \int\limits_{0}^{1} \rmd u_{2} \, \Bigg[ \Vert \vec{\Delta}_{\parallel} \Vert \, e^{-\frac{\Vert \vec{\Delta}_{\parallel} \Vert}{\ell_{r}}} \\
\nonumber & \left(1- \frac{4 \nu^{2} \Vert \vec{\Delta}_{\parallel} \Vert^{2}}{3! \Li^{2}}\right) - \Vert \vec{\Delta}_{\perp} \Vert \, e^{-\frac{\Vert \vec{\Delta}_{\perp} \Vert}{\ell_{r}}}  \left(1- \frac{4 \nu^{2} \Vert \vec{\Delta}_{\perp} \Vert^{2}}{3! \Li^{2}}\right) \Bigg] \\
& \left[1-\frac{1}{2!}\left(2 \nu \left( 1 - u_{1} \right) \right)^{2} - \frac{1}{2!} \left(2 \nu \left( 1 - u_{2} \right) \right)^{2} +\text{O}(\nu^{4}) \right] .
\end{align}
It is evident that $S_{\textsc{c}}(\nu=0)$ is non-zero (and also obviously positive). While it is also evident that this does not increase proportional to $\nu^{2}$ as in the inverse case, finding an analytical expression for the  above integral is not possible. We can, however, resort to numerically finding the logarithmic derivative of the PSD, as before. However, due to the dependence of $S_{\textsc{c}}(\nu)$ on the ratio $\rat=\ell_{r}/\Li$, the logarithmic derivative also changes with the choice of $\rat$. We find that the logarithmic derivative $S'_{\textsc{c}}(\nu)/S_{\textsc{c}}(\nu)\approx -8 \nu/\pi$ for $\rat=0.01$. This shows that $S_{\textsc{c}}(\nu \ll 1)\propto e^{-4 \nu^{2}/\pi}$ in this particular case. However, this fit changes significantly for different values of $\rat$.  

In the high-frequency limit, the sinc function already indicates a decay with increase in $\nu$. A full analytical expression cannot be obtained. However, we ascertain the high-frequency behaviour as follows. We know that if an expression is a sum of multiple terms with $\nu^{-q}$ ($q>0$) in the limit $\nu \gg 1$, the term most dominant is the one with the smallest $q$ value. So if we identify such a dominant term without solving the full PSD, we could still obtain the behaviour of the PSD at large $\nu$. Using the fact that $\Vert\vec{\Delta}_{\parallel}\Vert\leqslant\Vert\vec{\Delta}_{\perp}\Vert$ by geometry, we know that the terms involving $e^{-\frac{\Vert \vec{\Delta}_{\parallel} \Vert}{\ell_{r}}}$ would be the dominant terms, when computing the PSD. We consider one such term involved. 
\begin{align}
&\nonumber \frac{1}{\pi \ell_{r} } \int\limits_{0}^{1} \rmd u_{1}  \int\limits_{0}^{1} \rmd u_{2} \, \cos\left(2 \nu \left( u_{2} - u_{1} \right) \right)\\
& \Vert \vec{\Delta}_{\parallel} \Vert \, e^{-\frac{\Vert \vec{\Delta}_{\parallel} \Vert}{\ell_{r}}} \, \text{sinc}\left( 2 \nu \Vert \vec{\Delta}_{\parallel} \Vert /\Li \right).
\end{align}
Using $\Vert \vec{\Delta}_{\parallel} \Vert= \Li |u_{2}-u_{1}|$, we find that this simplies to
\begin{align}
&\nonumber  4 \frac{\ell_{r}  \Li}{\pi} \left(\frac{-2 \sqrt{2} \ell_{r} \Li}{\left(\Li^2+16 \ell_{r}^2 \nu^2 \right)^2}+\frac{1}{ \left(\Li^2+16 \ell_{r}^2 \nu^2 \right)}\right)\\
&\nonumber+\sqrt{2} \frac{\ell_{r}  \Li}{\pi} e^{-\frac{\Li}{\sqrt{2} \ell_{r}}} \bigg[\frac{8 \ell_{r} \Li}{\left(\Li^2+16 \ell_{r}^2 \nu^2 \right)^2} \cos\left(2 \sqrt{2} \nu \right)\\
&+\frac{\left(\Li^2-16 \ell_{r}^2 \nu^2 \right)}{\nu \left(\Li^2+16 \ell_{r}^2 \nu^2 \right)^2} \sin\left(2 \sqrt{2} \nu \right)\bigg]
\end{align}
We can see that the dominant trend in the above term is $\left(\Li^2+16 \ell_{r}^2 \nu^2 \right)^{-1}$. This tallies with the numerical evaluation of the PSD and is illustrated in Fig. 2 (c) in the paper.

\subsubsection{For $\rhoest$}
The PSD expression does not lend itself to analytic simplifications even in the limit cases. 

As illustrated in the case of $\rhoes$, here too we expect the behaviour at the low-frequency limit to be an exponential decay. However, obtaining the logarithmic derivative numerically poses challenges and would ultimately be dependent on the ratio $\rat$. So we forego the exercise in this case.

The high-frequency limit is more interesting. Though the higher limit is not analytically tractable, we obtain a few helpful pointers. It is known that the Fourier cosine transform of $e^{\alpha \sqrt{\varphi^{2}+\beta^{2}}}$ ($\text{Re}(\alpha)>0$, $\text{Re}(\beta)>0$) is given in terms of the modified Bessel function of the second kind $K_{1}(\beta \sqrt{\nu^{2}+\alpha^{2}})$ as follows. 
\begin{align}
\nonumber \int_{0}^{\infty} \rmd \varphi e^{\alpha \sqrt{\varphi^{2}+\beta^{2}}} \cos(\varphi \nu) &=  \alpha \beta (\nu^{2}+\alpha^{2})^{-1/2} \\
&K_{1}(\beta \sqrt{\nu^{2}+\alpha^{2}}) \label{eqn:FTBessel}
\end{align}
Notice the similarity of the LHS of Eq. \eref{eqn:FTBessel} to the RHS of Eq. \eref{eqn:TransCorrExp2}. Though this does not directly apply here, we expect the limiting behaviour to have some similarity in the gross feature, i.e., for instance, we expect it to tend as $e^{a \nu}/\nu^{b}$. A numerical fit yields $S_{\textsc{c}}(\nu)\approx 0.03 e^{-0.01} \nu^{-0.3}$ in the limit $\nu \gg 1$. Considering that the arguments used to obtain this fit are not rigorous, we agree that this need not be the correct behaviour. However, we find that the trend is distinctly different from that corresponding to $\rhoes$ or other correlation functions. This $S_{\textsc{c}}(\nu)$ corresponding to $\rhoest$ is, therefore, still \textit{distinguishable} from those corresponding to other $\rho$'s, even if we cannot identify an analytical limiting behaviour. 

\section{VII. Response function approach: Obtaining $\widetilde{\rho}$}

Using Eqs. \eref{eqn:2pt} and \eref{eqn:CorrTrans}, we find
\begin{align}
    \widetilde{\rhois}(\omega_{1},\kvec)= \frac{\ell_{r}}{ (2 \pi)^{2} \omega_{1} |\vec{k}_{1}|} \left[ \delta\left(|\vec{k}_{1}|-\frac{\omega_{1}}{c}\right) - \delta\left(|\vec{k}_{1}|+\frac{\omega_{1}}{c}\right) \right], \label{eqn:ZurekWave}
\end{align}
and
\begin{align}
    \nonumber \widetilde{\rhoes}(\omega_{1},\kvec)= &\frac{\ell_{r}^{2}}{ (2 \pi)^{2} \omega_{1} |\vec{k}_{1}|} \left[ \frac{1-\ell_{r}^{2} \left(|\vec{k}_{1}|-\frac{\omega_{1}}{c}\right)^{2}}{4\left(1+\ell_{r}^{2} \left(|\vec{k}_{1}|-\frac{\omega_{1}}{c}\right)^{2}\right)^{2}} \right.\\
    &- \left. \frac{1-\ell_{r}^{2} \left(|\vec{k}_{1}|+\frac{\omega_{1}}{c}\right)^{2}}{4\left(1+\ell_{r}^{2} \left(|\vec{k}_{1}|+\frac{\omega_{1}}{c}\right)^{2}\right)^{2}} \right].
\end{align}

It is evident from Eq. \eref{eqn:ZurekWave} that $\rhois$ implicitly assumes the wave equation when condsidering the SFs, which becomes apparent in $\widetilde{\rhois}$. This also conforms with the Pixellon model where the Pixellon is assumed to satisfy the wave equation~\cite{geontropy2023}. The transformed $\widetilde{\rhoes}$ is presented for contrast. As we have considered only isotropic models, we also find the transformed correlation functions depend only on the magnitude of $\vec{k}_{1}$.

\section{VIII. LIGO: Behaviour of PSD}

As in Secs. V and VI, we examine the PSD of the output of LIGO, corresponding to two different classes of the correlation function, especially at the low- and high-frequency limits.  It is evident from Eq. \eref{eq:RFaLIGO} that the behaviour of the PSD of the output of LIGO can be examined by examining the behaviour of the PSD in the case of the MLI without arm cavities and the Fabry-P\'erot cavity response, given by Eq. \eref{eq:RFcav}.

\textit{Low-frequency limit:} In the limit $\nu\ll1$, the Fabry-P\'erot cavity response in Eq. \eref{eqn:FPgain} is clearly shown to be inversely proportional to $\nu^{2}$ and as $\nu\to0$, this saturates to $T_{\textsc{m}}^{4} \left( \frac{1}{1- \sqrt{R_{\textsc{m}}}} \right)^{6}$, instead of diverging. 

In the case of the correlation function $\rhois$, we know from Sec. VI that the PSD of an MLI without arm cavities is directly proportional to $\nu^{2}$. This implies that when $\nu\ll 1$, the PSD of the output of LIGO is constant with respect to $\nu$. This is because the frequency dependence of the Michelson interferometric response cancels that of the Fabry-P\'erot cavity response. However, as $\nu\to0$, the PSD of LIGO becomes directly proportional to $\nu^{2}$, because of the saturation in the Fabry-P\'erot cavity response. 

In the case of $\rhoes$, it is proportional to an exponential factor which is almost constant in this limit. This implies that the low-frequency behaviour of the Fabry-P\'erot cavity response is identical to that of the PSD of LIGO.

\textit{High-Frequency limit:} In the limit $\nu \gg 1$, the Fabry-P\'erot cavity response maximises to $T_{\textsc{m}}^{4} \left( \frac{1}{1- \sqrt{R_{\textsc{m}}}} \right)^{6}$ at every $4\nu=2 m \pi$ ($m\in\{1,2,3,\dots\}$), resulting in peaks at every $f=m\fsr$. The trend of the MLI response is also carried forward. For instance, this is evident from top panel of Fig. 4 in the paper, where the troughs between the peaks fall as $1/\nu$, which is the behaviour corresponding to the MLI for $\rhois$. This is also illustrated in the bottom panel of Fig. 4 in the paper. Here the response remains flat in the range of $\nu$ considered due to the significantly smaller $r$. This behaviour is identical to the PSD correponding to an MLI of the same arm length as LIGO, but with no arm cavities. In both cases, the peaks with the gain $T_{\textsc{m}}^{4} \left( \frac{1}{1- \sqrt{R_{\textsc{m}}}} \right)^{6}$ render a PSD of the arm strain in LIGO that is significantly larger than the same in table-top interferometers such as QUEST.


\bibliography{references}